\documentclass[sigconf]{acmart}

\usepackage{graphicx}
\usepackage{comment}
\usepackage{amsmath} % define this before the line numbering.
\usepackage{color}
\usepackage{microtype}
\usepackage{subfigure}
\usepackage{booktabs} % for professional tables
\usepackage{multirow}
\usepackage{diagbox}
\usepackage{appendix}
\usepackage[linesnumbered, lined, boxed, commentsnumbered, ruled, vlined]{algorithm2e}
\usepackage{multicol}
\usepackage{balance}

\newcommand{\eat}[1]{}

\newcommand{\smalltitle}[1]{ \vspace{1mm}{\noindent\textbf{#1.}\hspace{1mm}}}
\newcommand{\add}[1]{{#1}}
\newcommand{\changecolor}[1]{}

\def\L{\mathcal{L}}

\def\ours{PoE\xspace}

\copyrightyear{2021}
\acmYear{2021}
\setcopyright{acmcopyright}\acmConference[SIGMOD '21]{Proceedings of the 2021 International Conference on Management of Data}{June 20--25, 2021}{Virtual Event, China}
\acmBooktitle{Proceedings of the 2021 International Conference on Management of Data (SIGMOD '21), June 20--25, 2021, Virtual Event, China}
\acmPrice{15.00}
\acmDOI{10.1145/3448016.3457326}
\acmISBN{978-1-4503-8343-1/21/06}
\begin{document}
\fancyhead{}

\title{Pool of Experts: Realtime Querying Specialized Knowledge in Massive Neural Networks}

%
% The "author" command and its associated commands are used to define
% the authors and their affiliations.
% Of note is the shared affiliation of the first two authors, and the
% "authornote" and "authornotemark" commands
% used to denote shared contribution to the research.
\author{Hakbin Kim}
\email{qlsgkrrla@naver.com}
\affiliation{%
  \institution{Inha University}
  \city{Incheon}
  \country{South Korea}
}
\author{Dong-Wan Choi}
\email{dchoi@inha.ac.kr}
\affiliation{%
  \institution{Inha University}
  \city{Incheon}
  \country{South Korea}
}

% \author{Anonymous}

%
% By default, the full list of authors will be used in the page
% headers. Often, this list is too long, and will overlap
% other information printed in the page headers. This command allows
% the author to define a more concise list
% of authors' names for this purpose.
% \renewcommand{\shortauthors}{Hakbin Kim and Dong-Wan Choi}

\begin{abstract}
In spite of the great success of deep learning technologies, training and delivery of a practically serviceable model is still a highly time-consuming process. Furthermore, a resulting model is usually too generic and heavyweight, and hence essentially goes through another expensive model compression phase to fit in a resource-limited device like embedded systems. Inspired by the fact that a machine learning task specifically requested by mobile users is often much simpler than it is supported by a massive generic model, this paper proposes a framework, called \textit{Pool of Experts} (PoE), that instantly builds a lightweight and task-specific model without any training process. For a realtime model querying service, PoE first extracts a pool of primitive components, called \textit{experts}, from a well-trained and sufficiently generic network by exploiting a novel \textit{conditional knowledge distillation} method, and then performs our \textit{train-free knowledge consolidation} to quickly combine necessary experts into a lightweight network for a target task. Thanks to this train-free property, in our thorough empirical study, PoE can build a fairly accurate yet compact model in a realtime manner, whereas it takes a few minutes per query for the other training methods to achieve a similar level of the accuracy.

\keywords{Knowledge distillation \and Model compression \and Model specialization \and Model unification}
\end{abstract}

\begin{CCSXML}
<ccs2012>
<concept>
<concept_id>10010147.10010257</concept_id>
<concept_desc>Computing methodologies~Machine learning</concept_desc>
<concept_significance>500</concept_significance>
</concept>
<concept>
<concept_id>10002951.10003227.10003351</concept_id>
<concept_desc>Information systems~Data mining</concept_desc>
<concept_significance>500</concept_significance>
</concept>
<concept>
<concept_id>10002951.10003227.10003245</concept_id>
<concept_desc>Information systems~Mobile information processing systems</concept_desc>
<concept_significance>300</concept_significance>
</concept>
</ccs2012>
\end{CCSXML}

\ccsdesc[500]{Computing methodologies~Machine learning}
\ccsdesc[500]{Information systems~Data mining}
\ccsdesc[300]{Information systems~Mobile information processing systems}

\keywords{Lightweight Neural Networks, Knowledge Distillation, Model Specialization, Model Compression}


% \include{coverletter} % to be deleted in the CRC version

\maketitle

\section{Introduction} \label{sec:intro}
%  Hook
Imagine a realtime \textit{AI-as-a-Service} (\textit{AIaaS}) system that can instantly deliver \textit{resource-efficient} models for any on-demand tasks \add{to multiple users who can be non-expertise in AI}. For example, a mobile \add{user} should quickly adapt to a dynamically changing environment \add{(e.g., entering a restaurant in an animal theme park and returning to see animals having lunch)} preferably without continually training a large generic model due to the resource constraints. In this case, such a realtime AIaaS system would be an adequate solution so that a user can immediately be given a properly functioning model by the system without any training overhead. Although AIaaS is getting popular to the point that global vendors have initiated code-free machine learning platforms, it still takes a while to train a specific neural network that achieves a practical level of the accuracy even with the help of transfer learning from a massive pretrained model, often referred to as \textit{oracle}. To realize a realtime AIaaS system that immediately builds a task-specific and lightweight model, would it be possible to preprocess such an oracle neural network so that any queried knowledge is efficiently extracted with no training at all?

% KD is good for lightweight model construction..
The fundamental problem of extracting knowledge from a pretrained model has been well studied in the context of \textit{knowledge distillation} (KD) \cite{DBLP:journals/corr/HintonVD15}. The standard KD method is basically intended for transferring the entire knowledge from a large teacher model to a small student model, and is reported to be fairly effective in reducing the model size as well as preserving the high accuracy \cite{DBLP:journals/corr/HintonVD15,DBLP:journals/corr/RomeroBKCGB14,DBLP:conf/iclr/ZagoruykoK17,DBLP:conf/cvpr/YimJBK17}.
% Drawbacks of standard KD in model compression...
However, if a target task is much simpler than the one supported by the oracle, which is pretty common in many IoT applications, the direct use of standard KD would be neither efficient nor effective as a vast knowledge irrelevant to the target task will also be transferred. Consider a mobile image recognition system (e.g., Google Lens\footnote{https://lens.google.com/}), where users want to be informed what kind of objects that they see in a particular place such as foods in a restaurant or animals in a zoo. If we try to compress a fully generic image classifier trained for millions of classes for this application, we may end up with either a lightweight but less accurate model or a still too heavyweight model to fit in a mobile device, not to mention that model compression itself is an expensive process.

% Model specialization methods focusing on the above problems, but they have drawbacks...
\textit{Model specialization} methods have been utilized in order to address such a tradeoff between model size and accuracy by training a specialized model for a target task \cite{DBLP:conf/mobisys/HanSPAWK16,DBLP:conf/cvpr/ShenHPK17,DBLP:journals/pvldb/KangEABZ17,DBLP:conf/osdi/HsiehABVBPGM18}. Although these techniques can help to build a smaller and accurate model, we should somehow train the model from scratch whenever a user requests a new task, which is far from the aforementioned realtime AIaaS system, \add{not to mentioned that the user has to provide a task-specific dataset to be trained. Alternatively, we can pre-train the complete set of specialized models for all the possible combinations of tasks, which however would require a huge amount of storage and training overhead.}

% Another serious drawback of model specialization lies in the fact that we always need a task-specific dataset to be trained, which is however not expected in practice as it is hard to prepare the dataset in advance for every task on request.

\begin{figure}[t!]
    \centering
    % preprocessing phase
    \subfigure[\label{fig:framework:a}Preprocessing Phase: Knowledge Extraction and Decomposition]{\includegraphics[width=0.9\columnwidth]{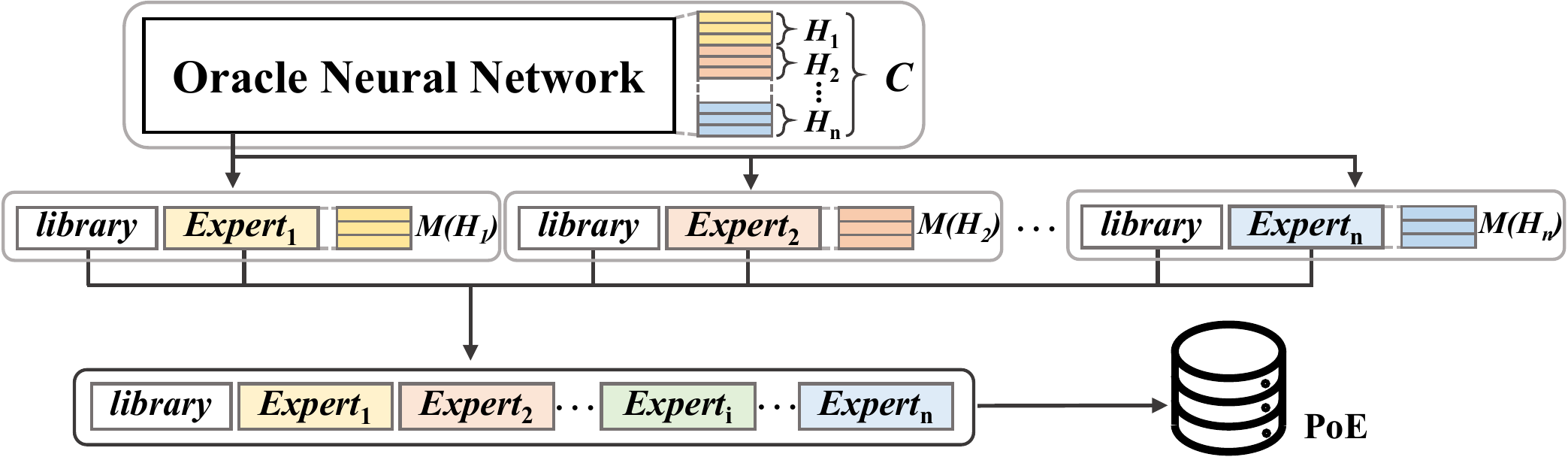}}
    % service phase
    \subfigure[\label{fig:framework:b}Service Phase: Knowledge
    Consolidation]{\includegraphics[width=0.9\columnwidth]{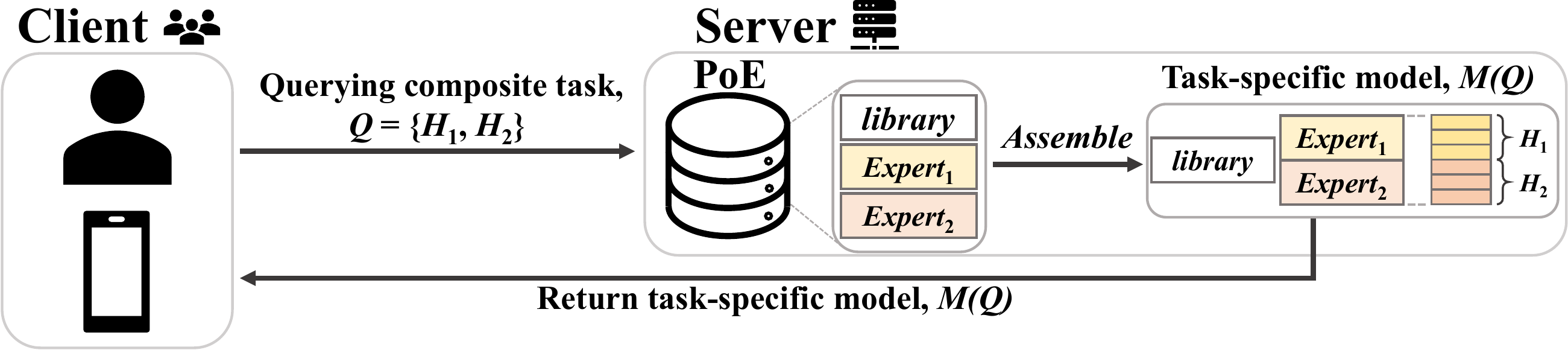}} 
    \vspace{-4mm}
    \caption{Overall procedure of Pool of Experts (\ours)}
    \label{fig:framework}
    \vspace{-4mm}
\end{figure}

% Our framework...
Toward a \textit{realtime model querying system}, this paper proposes a novel framework, called \textit{Pool of Experts} (PoE), that can instantly build a specialized and lightweight model for a target queried task \add{without storing all the possible specialized models}. The main strategy of PoE is to extract a set of composable components from a pretrained and sufficiently generic model (i.e., oracle) and combine some of them to synthesize an efficient model for the requested task, as illustrated in Figure \ref{fig:framework}. More specifically, in the preprocessing phase, PoE first constructs a pool of specialized and primitive components, called \textit{experts}, along with a shared component, called \textit{library}, from the oracle network. Then, in the service phase, when a user actually asks a model for a specific task, PoE quickly combines the library part and a minimal set of experts required for the target task without any training process, thereby building the queried task-specific model in a realtime fashion.

% Challenges (confidence)
Both preprocessing and service phases are collectively challenging. First, it is not straightforward to extract a pool of experts particularly in a way that they can readily be composed for the service phase. One possible way is to train each expert by transfer learning on its task-specific data from a smaller version of oracle compressed by standard KD. Unfortunately, to be analyzed in Section \ref{sec:experiment:special}, this approach can produce erroneously confident experts that can make an incorrect inference with high confidence even for a class it has never seen. Such an \textit{overconfident} expert is problematic when assembling multiple experts for a unified model. Furthermore, even if we extract properly confident experts, it appears impossible to combine them into a single model without any training process in that most of the existing works on merging neural networks include some kinds of training phase \cite{VongkulbhisalVS19,DBLP:journals/corr/abs-2003-03701}.

PoE addresses the challenges above by the following two strategies. For the preprocessing phase, we devise a novel \textit{conditional knowledge distillation} (CKD) method, which not only effectively extracts the specialized knowledge for each expert but also makes each expert properly confident to tell which classes are unknown to it. Also, we propose \textit{train-free knowledge consolidation} that exploits a \textit{logit concatenation} scheme in order to combine multiple experts into a single model in realtime. This causes another difficult challenge, called the \textit{logit scale} problem, where the issue is how to normalize multiple logits in arbitrary scales. We also solve this sub-problem of independent interest by defining a new regularization term in our CKD loss function.

% The scope of this paper & experimental result summary
Even though our PoE framework can be applicable to any oracle models whose task can be decomposed into a set of primitive ones, this paper focuses on image classification, in which there are various benchmark models generic and accurate enough to be called oracles. Through extensive experiments using CIFAR-100 \cite{krizhevsky2009learning} and Tiny-ImageNet \cite{le2015tiny} datasets, we show that PoE can instantly build any queried image classifier with a practical level of the accuracy yet \emph{two orders of magnitude} less parameters than oracle without any training process.

% Contributions
In summary, we make the following contributions in this paper: (1) We present \textit{Pool of Experts} (PoE), which is a novel framework toward a realtime model querying system, where a lightweight and task-specific classifier is instantly extracted from a massive generic neural network in response to a particular queried task. This is the first study on realtime processing of model queries over massive neural networks. (2) We devise \textit{conditional knowledge distillation} (CKD) in order to extract only the specialized knowledge from a pretrained neural network. (3) We propose \textit{train-free knowledge consolidation} so that we can merge multiple experts into a single unified model without any training process. (4) A thorough experimental study is performed in image classification. Experimental results show that PoE is able to build a fairly accurate and resource-efficient model in a realtime fashion, compared to the other training methods taking at least a few hundreds of seconds for each query to obtain even less accurate models.

\section{Related Works} \label{sec:related}
In this section, we consider the following major branches of works strongly related to ours: neural network compression, model specialization, and model unification.

\smalltitle{\add{Neural network compression}}
\add{Motivated by intelligent mobile applications, the problem of compressing overparameterized neural networks has attracted many researchers in the deep learning community. We can arguably categorize numerous neural network compression techniques into three major branches, which are \textit{knowledge distillation} (KD) \cite{DBLP:conf/kdd/BucilaCN06,DBLP:journals/corr/HintonVD15}, \textit{quantization} \cite{DBLP:conf/eccv/RastegariORF16,DBLP:conf/iclr/ZhouYGXC17,DBLP:conf/cvpr/JacobKCZTHAK18}, and \textit{network pruning} \cite{DBLP:conf/nips/HanPTD15, DBLP:conf/iccv/LuoWL17,DBLP:conf/cvpr/HeLWHY19}. These three schemes are often considered to be orthogonal to each other and therefore collectively used to improve the final performance. This paper focuses on KD, which is complementary to any methods on quantization and network pruning.}

The idea of transferring the knowledge from a larger model to a smaller model is initiated in \cite{DBLP:conf/kdd/BucilaCN06}, and then it is further generalized with the name KD in \cite{DBLP:journals/corr/HintonVD15}. In KD, the output probabilities of a teacher model is softened by a hyperparameter called \textit{temperature}, and a student model can learn the teacher's knowledge by trying to mimic this softened label. Later studies \cite{DBLP:journals/corr/RomeroBKCGB14,DBLP:conf/iclr/ZagoruykoK17,DBLP:conf/cvpr/YimJBK17} based on the KD method have been continued, but they all focus on how to improve the effectiveness of distilling the entire knowledge of teacher so that the overall accuracy gets better. The main difference with these methods of our \ours framework is that our CKD method aims to extract only the specialized knowledge from the teacher model and deliver it to the student model.

\smalltitle{Model specialization}
Particularly in the database community, model specialization methods are being employed to reduce the inference time in image or video querying systems \cite{DBLP:conf/cvpr/ShenHPK17,DBLP:journals/pvldb/KangEABZ17,DBLP:conf/osdi/HsiehABVBPGM18,DBLP:conf/iccv/MullapudiCZRF19,DBLP:conf/mobisys/HanSPAWK16,DBLP:journals/pvldb/KangBZ19,DBLP:conf/icde/KoudasLX20}. They are basically aimed at building a small specialized neural network than a large oracle model, which is similar to the purpose of this paper. To this end, \textit{MCDNN} \cite{DBLP:conf/mobisys/HanSPAWK16} first compresses a generic model through various model compression techniques \cite{DBLP:conf/icml/ChenWTWC15,DBLP:conf/nips/HanPTD15,DBLP:conf/icassp/XueLYSG14}, and thereby construct smaller versions of oracles. Then, it re-trains only the output layer of these smaller oracles to specialize them for a given target task. In \textit{NoScope} and its extended version \cite{DBLP:journals/pvldb/KangEABZ17,DBLP:journals/pvldb/KangBZ19}, multiple smaller architectures are trained as candidate specialized models, and the one with a reasonably good accuracy beyond a predefined threshold is selected. In all these methods, we should carry out either additional training or training from scratch to build a specialized model whenever a new target task is given. However, \ours can synthesize the specialized model for any queried task in realtime because there is no further training process after extracting experts from oracle in the preprocessing phase.

% Subsequently, only the data for specific tasks that want specialization from the data used to train the generalized model are extracted and the output layer of the compressed generalization model is retrained to create the specialized model. \cite{DBLP:journals/pvldb/KangEABZ17,DBLP:journals/pvldb/KangBZ19} receives sample data for specific tasks that want to specialize by input, and a generalized model that is pre-trained for the entire task, and uses generalized models to generate an answer label for sample data. Afterwards, various small neural network architectures will be placed as candidates for specialized models, and after training through sample data, the optimal neural network will be selected. 

\smalltitle{Model unification}
The problem of unifying different classifiers has long been studied in the machine learning community. A typically known way is the ensemble method \cite{DBLP:journals/pami/KittlerHDM98,DBLP:books/wi/04/K2004}, where we combine multiple models by voting or averaging over their outputs. However, ensemble methods assume that every model is built for the same task, and therefore are not applicable to merging multiple specialized models like experts of \ours. To the best of our knowledge, in our framework, the method most closely related to the scenario of merging experts is \textit{UHC} \cite{VongkulbhisalVS19}. The UHC method intends to merge heterogeneous neural networks that have been independently trained for different classes by distilling the knowledge from multiple teachers into a single student model. This method can be used to merge experts in our framework as well. Unfortunately, to be experimentally shown in Section \ref{sec:experiments:service}, UHC does not only require a non-trivial time of training, but also generates models even less accurate than the task-specific models built by \ours that needs no training at all. Moreover, we observe that UHC is greatly influenced by how each teacher model is trained. If all the teachers are disjoint in their classes and separately trained from scratch, models combined by UHC turn out to get significantly worse due to the lack of the knowledge across experts. Similar to UHC, \textit{DMC} \cite{DBLP:conf/wacv/ZhangZGLTHZK20} extends the standard KD method to combine two disjoint models by using its proposed \textit{double distillation} method. For the problem of continual learning \cite{ParisiKPKW19}, they focus on how to combine the two models, namely the previously trained old model and the one trained for a new task. It would be possible to extend DMC to merge multiple models, and we believe such an extension is pretty close to how UHC works. Thus, DMC can be seen as a special case of UHC in the context of the merging functionality, which still needs a further training phase and would suffer from the same issue as UHC when multiple models have to be merged.

% \subsection{Branched CNNs}
% There are several studies\cite{DBLP:conf/iccv/YanZPJDDY15,DBLP:conf/eccv/AhmedBT16,DBLP:journals/ijon/WangWW18} related to Branched CNN. The structure of \cite{DBLP:conf/eccv/AhmedBT16} has a shared network part at the front and consists of expert branches at the back. The structure of \cite{DBLP:conf/iccv/YanZPJDDY15} consists of coarse components trained over all classes and fine components trained over subsets of classes. Coarse components and fine components also share shared layers, who play the role of feature extractor. The above studies aim to use tree-structure to distinguish subsets of classes that are difficult to distinguish from each other. Our framework(PoE) constructs the shared feature extractor and experts in a tree-structure similar to the above papers. However, our framework aims to reduce the number of FLOPs of specialized models by using tree-structure to break down Oracle into multiple components.
\section{Problem Definition} \label{sec:prob}
In this section, we formally define our proposed problem toward a realtime model delivery system, particularly focusing on image classification.
\begin{itemize}
    \item Given a set $C$ of classes, an \textit{oracle neural network}, denoted by $M(C)$, is a pretrained classification model for $C$.
    \item $C$ can be divided into $n$ \textit{primitive tasks} for $n \leq |C|$, each of which is a subset $H_i$ of $C$, that is, $C = \bigcup_{i=1}^{n} H_i$, such that each $H_i$ is sufficiently fine-grained and therefore does not need to be further decomposed. For instance, if there is a class hierarchy for $C$, each primitive task can correspond to a super class at a properly low level of the hierarchy.
    \item A \textit{composite task} $Q$ is a union set of multiple primitive tasks, where the number of primitive tasks in $Q$ is denoted by $n(Q)$. 
    \item Given a composite task $Q \subseteq C$, the \textit{task-specific model} for $Q$, denoted by $M(Q)$, is a specialized model that can recognize any images of a class in $Q$.
\end{itemize}

\smalltitle{Problem statement} Given $M(C)$, our problem aims to preprocess $M(C)$ so that we can build the task-specific model $M(Q)$ for any queried composite task $Q$ such that: (1) the size of $M(Q)$ is much smaller than that of $M(C)$ when $|Q| \ll |C|$, (2) the accuracy of $M(Q)$ is comparable to that of $M(C)$, and (3) the model construction process runs as fast as possible.

\section{Realtime Querying Specialized Knowledge} \label{sec:main}

% Overview
This section describes our framework, called \textit{Pool of Experts} (PoE), that can instantly build any task-specific and lightweight model for a composite task on request. The PoE framework works in the following two phases. In the preprocessing phase, we decompose a given oracle network into a pool of specialized components, called \textit{experts}, which constitute our PoE framework, and then PoE quickly assembles a group of necessary experts to synthesize the task-specific model for a queried composite task in the service phase.

% How to build Pool of Experts?
\subsection{Preprocessing Phase: Knowledge Extraction and Decomposition} \label{sec:main:preprocess}

%%% high conf reason figure
\begin{figure*}[t!]
    \centering
    \vspace{-4mm}
    \subfigure[Training $M(H_1)$ by the cross entropy loss with hard targets]{\hspace{7mm}\includegraphics[height=35mm]{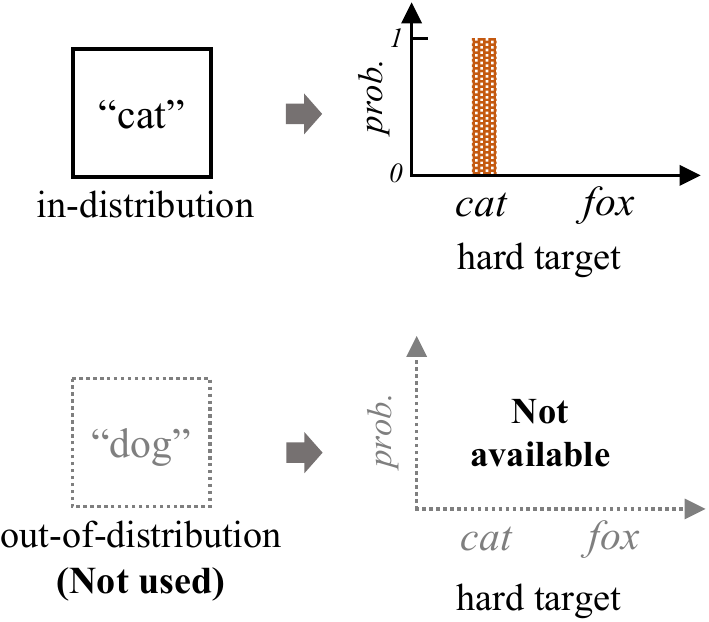}\label{fig:high conf reason:hard label}\hspace{7mm}}
    \subfigure[Training $M(H_1)$ by $\L_{soft}$ with soft targets]{\includegraphics[height=31mm]{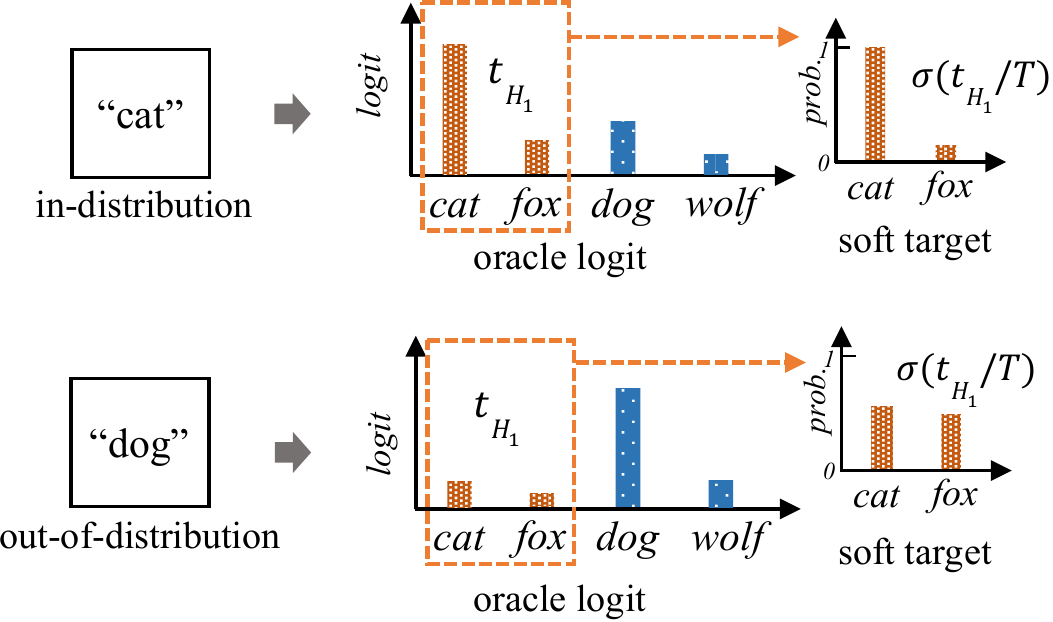}\label{fig:high conf reason:soft target}}
    \subfigure[Inference with out-of-distribution image $dog$ on $M(H_1)$ trained by either the cross entropy loss or $\L_{soft}$]{\raisebox{0.34\height}{\includegraphics[height=21mm]{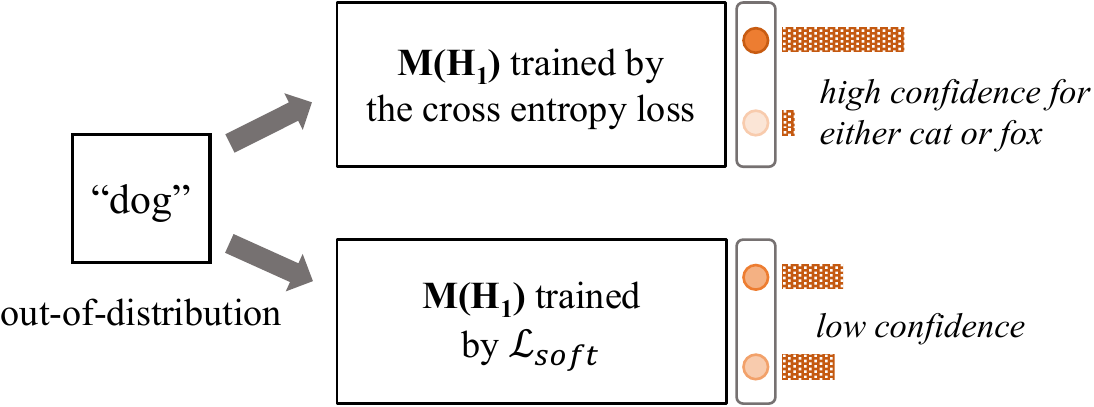}}\label{fig:high conf reason:inference}}
    \vspace{-4mm}
	\caption{Difference between the cross entropy loss with hard targets and our $\L_{soft}$ loss with soft targets when training experts for primitive task $H_1 = \{cat, fox\}$}
	\label{fig:high conf reason}
	\vspace{-4mm}
\end{figure*}

\smalltitle{Library extraction}
In order for a resulting task-specific model to be as lightweight as possible, one essential property of PoE is that the size of each expert must be also tiny. Otherwise, the total size of assembled experts could be non-trivially large particularly when a composite task contains multiple primitives. Our strategy is to extract the common knowledge from oracle, called \textit{library}, to be shared by all the experts. The library component should not only be small enough, but also contain common features for all the primitive tasks of oracle. To this end, we perform standard KD \cite{DBLP:journals/corr/HintonVD15} to distill from oracle to a smaller model that still covers all the primitive tasks by the following loss function:
\begin{equation}
\mathcal{L}_{KD} = D_{KL}\left(\sigma\left(\mathbf{t}/T\right)\parallel \sigma\left(\mathbf{s}/T\right)\right), 
\label{KD loss}
\end{equation}
where $\mathbf{t}/T$ and $\textbf{s}/T$ are softened logits with a temperature parameter $T$ and output logits $\mathbf{t}$ and $\textbf{s}$ returned by oracle and the smaller target model, respectively. For two probability distributions $P$ and $Q$, $D_{KL}(P \parallel Q)$ represents the well-known \textit{Kullback–Leibler divergence} defined as $\sum_{c \in C} P(c)\log{\frac{P(c)}{Q(c)}}$.

Once this KD process is done, we take the first $\ell$ layers from the trained target model to be the library component, where $\ell > 1$ is a hyperparameter that controls the tradeoff between the size of a task-specific model and its accuracy.

\smalltitle{Experts extraction}
Now that the library component is extracted, all we have to do is to extract $n$ experts from oracle, each of which represents the specialized knowledge for a primitive task $H_i \subset C$. To this end, this paper proposes \textit{conditional knowledge distillation} (CKD), which is based on standard KD but distills only the specialized knowledge of each $H_i$ to a target model by introducing two new loss terms to the CKD loss as follows:
\begin{equation} \label{CKD loss}
\mathcal{L}_{CKD} = \mathcal{L}_{soft}\; +\; \alpha\,\mathcal{L}_{scale},
\end{equation}
where $\alpha$ is a hyperparameter controlling the ratio of the two losses, namely $\mathcal{L}_{soft}$ and $\mathcal{L}_{scale}$. For each primitive task $H_i$, we basically distill the specialized knowledge acquired by $\mathcal{L}_{CKD}$ from oracle to its \textit{target expert model}, i.e., $M(H_i)$. As depicted in Figure \ref{fig:framework:a}, each $M(H_i)$ starts with the same library component and is attached with its own component to be taken as the expert for $H_i$. During the extraction process, we freeze the library component and update only the expert component. 

Let us now look into each of the two losses of $\L_{CKD}$. First, $\mathcal{L}_{soft}$ is a specialized version of $\mathcal{L}_{KD}$, which intends to extract only the knowledge essentially required for each primitive task $H_i$ by:
\begin{equation} \label{soft loss}
\mathcal{L}_{soft} = D_{KL}\left(\sigma\left(\mathbf{t}_{H_i}/T\right)\parallel \sigma\left(\mathbf{s}_{H_i}/T\right)\right),
\end{equation}
where $\mathbf{t}_{H_i}/T$ and $\mathbf{s}_{H_i}/T$ are the softened logits of oracle and a target expert model $M(H_i)$, respectively, similar to the definition of $\L_{KD}$ in Eq. (\ref{KD loss}). At this time, however, $\mathbf{t}_{H_i}$ represents not the full output logit $\mathbf{t}$ of oracle, but the \textit{sub-logit} that corresponds to only the classes in $H_{i}$. Thus, $\mathbf{t}_{H_i}$ as well as $\mathbf{s}_{H_i}$ has the length $|H_{i}|$.

% In order to obtain $\mathbf{s}_{H_i}$, for all $i \in [1, n]$, we freeze the library component and take the resulting logit after the forward propagation on $M(H_i)$ to be trained for each expert for $H_i$. 

% To this end, this paper proposes \textit{conditional knowledge distillation} (CKD), which is based on standard KD but distills only the specialized knowledge of each $H_i$ to a target model by introducing two new loss terms to the CKD loss as follows:
% \begin{equation} \label{CKD loss}
% \mathcal{L}_{CKD} = \mathcal{L}_{soft}\; +\; \alpha\,\mathcal{L}_{scale},
% \end{equation}
% where $\alpha$ is a hyperparameter controlling the ratio of the two losses, namely $\mathcal{L}_{soft}$ and $\mathcal{L}_{scale}$. For each primitive task $H_i$, we basically distill the specialized knowledge acquired by $\mathcal{L}_{CKD}$ from oracle to a \textit{target expert model}, i.e., $M(H_i)$. Each $M(H_i)$ starts with the same library component and is attached with its own component to be taken as the expert for $H_i$. During the training process, we freeze the library component and only the expert component is updated. 

% Each expert model $M(H_i)$ starts with the same library component and is attached with its own component to be taken as the expert for $H_i$. During the training process, we freeze the library component and only the expert component is updated. 

Through the loss $\mathcal{L}_{soft}$, each expert can learn only the specialized knowledge of its corresponding primitive task, which is likely to need much less capacity than it takes to learn the entire knowledge of oracle. To be experimentally confirmed in Section \ref{sec:experiment:special}, a specialized model trained by CKD is able to achieve an accuracy comparable to that of oracle with only about 150 times less parameters and 65 times less FLOPs, and shows a much higher accuracy than the same lightweight architecture trained by $\mathcal{L}_{KD}$.

The role of the second term $\mathcal{L}_{scale}$ is strongly related to our \textit{train-free knowledge consolidation} scheme, to be explained in Section \ref{sec:main:service}.

\begin{figure*}[t]
	\begin{minipage}[h]{\columnwidth}
    	\centering
    	\includegraphics[width=0.79\columnwidth]{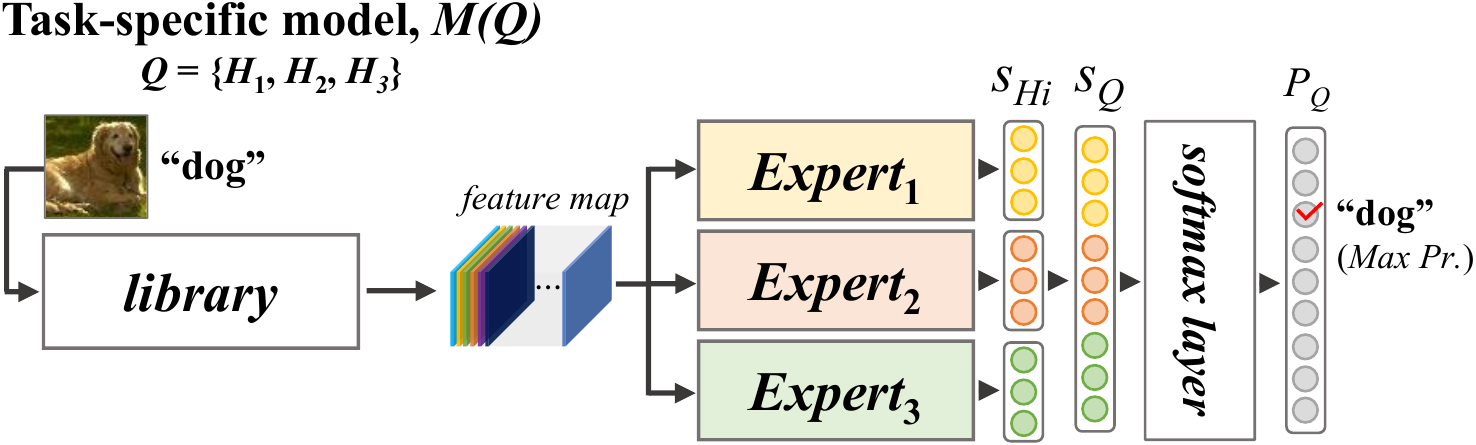}
    	\vspace{-3mm}
    	\caption{The branched architecture of a task-specific model}
    	\label{fig:branched architecture}
	\end{minipage}
    \hspace{4mm}
	\begin{minipage}[h]{\columnwidth}
    	\centering 
        \subfigure[\label{fig:logitprob:a}Properly confident experts, $M(H_1)$ and $M(H_2)$]{\includegraphics[height=16mm]{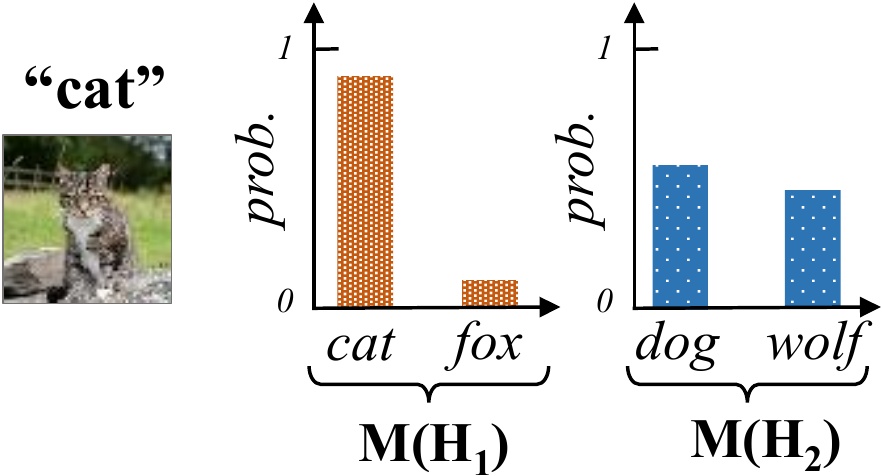}}
        \subfigure[\label{fig:logitprob:b}Logit scale problem on $M(Q)$]{\includegraphics[height=17mm]{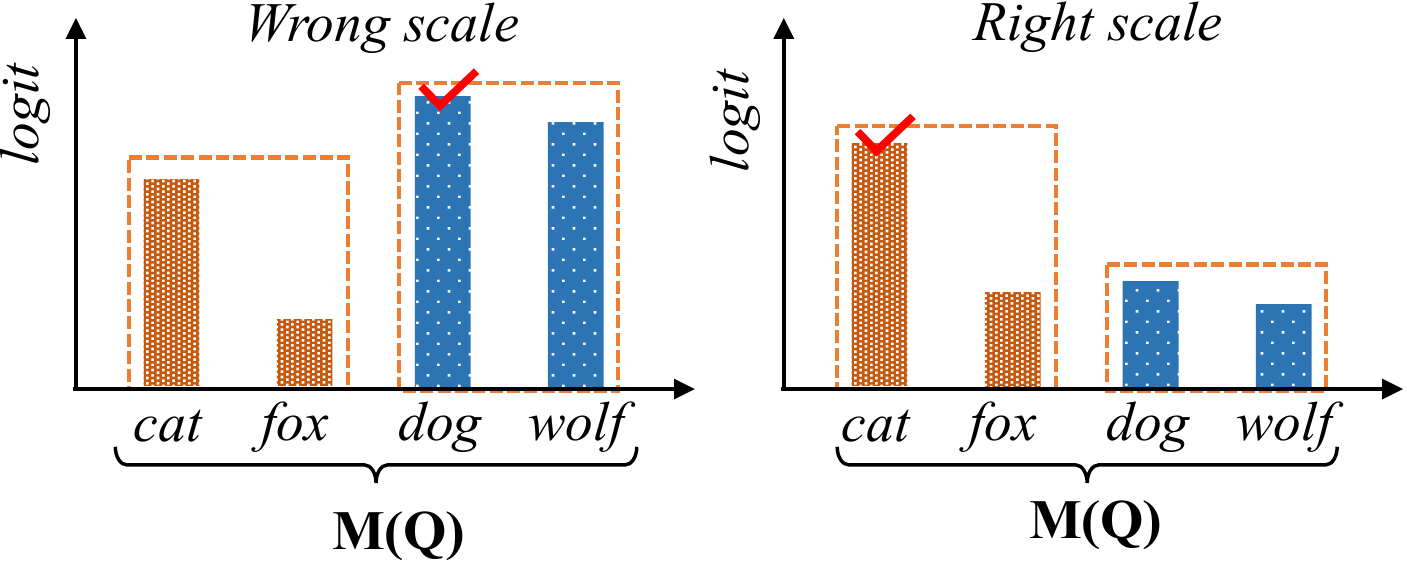}} 
        \vspace{-3mm}
    	\caption{An example to illustrate the logit scale problem, where $Q = H_1 \cup H_2$, $H_1 = \{cat, fox\}$, and $H_2 = \{dog, wolf\}$}
    	\label{fig:logitprob}
	\end{minipage}
	\vspace{-4mm}
\end{figure*}

\smalltitle{Avoiding overconfident experts}
As mentioned in the introduction, the main challenge of extracting experts lies in the fact that each expert should be in an easily composable form to be assembled with the other experts. For this purpose, the loss $\mathcal{L}_{soft}$ plays a key role by preventing erroneous experts that are too confident about classes unknown to them. When we assemble multiple experts to consolidate their knowledge, this property is particularly important because an overconfident expert can hurt the correct prediction made by a true expert who actually has the right knowledge. 

To illustrate, suppose we train the expert model $M(H_1)$ for a primitive task $H_1 = \{cat, fox\}$ as shown in Figure \ref{fig:high conf reason}. An ideally trained $M(H_1)$ should not be highly confident about \textit{out-of-distribution} images, which belong to irrelevant classes like $dog \notin H_1$. However, if we simply train $M(H_1)$ either from scratch or by transfer learning, then the resulting model can be \textit{overconfident} about out-of-distribution images as depicted by the upper $M(H_1)$ trained by the \textit{cross entropy} loss in Figure \ref{fig:high conf reason:inference}. The underlying reason is that we take only hard targets of \textit{in-distribution} samples when training from scratch by transfer learning with the cross entropy loss, and do not take into account out-of-distribution samples (see Figure \ref{fig:high conf reason:hard label}). For neural networks typically trained with only in-distribution samples, recent works \cite{DBLP:conf/iclr/HendrycksG17,DBLP:conf/cvpr/NguyenYC15} have reported that they tend to make high confidence even for completely irrelevant inputs. 

On the other hand, as illustrated in Figure \ref{fig:high conf reason:soft target}, when minimizing $\L_{soft}$ to distill the specialized knowledge for $H_1$ from oracle, we can extract the soft target $\sigma(\mathbf{t}_{H_1}/T)$ from out-of-distribution samples as well as in-distribution samples. For the samples irrelevant to $H_1$, oracle would output a lowly distributed sub-logit for $H_1$, and thereby the expert model can learn this knowledge of low confidence for out-of-distribution samples as well. In Section \ref{sec:experiment:special}, we will show through our experimental results that the model trained by $\mathcal{L}_{soft}$ has low confidence for out-of-distribution examples like the bottom $M(H_1)$ in Figure \ref{fig:high conf reason:inference}. 

\subsection{Service Phase: Train-Free Knowledge Consolidation} \label{sec:main:service}
\smalltitle{Knowledge consolidation by logit concatenation}
In the service phase, PoE assembles the library component and the multiple expert components required for a given composite task in realtime, and thereby builds a single task-specific model without any training process. Our main strategy for this train-free model generation is to organize the task-specific model to have multiple branches from library, one for each expert, yet with a single final layer where the knowledge is finally consolidated. More specifically, as shown in Figure \ref{fig:branched architecture}, we just put the library component at the front of the model, and make the input of all the necessary experts to be connected to library. Finally, the output logits of experts (i.e., $\mathbf{s}_{H_i}$) are \textit{concatenated} into a unified logit, denoted by $\mathbf{s}_{Q}$, to be an input of the final softmax function.

Somewhat surprisingly, this simple way of \textit{logit concatenation} works very well for our problem of building a task-specific model. To be experimentally shown in Section \ref{sec:experiments:service}, however, this does not necessarily imply that specialized models distilled by using only $\mathcal{L}_{soft}$ can be easily unified by logit concatenation even though $\mathcal{L}_{soft}$ effectively addresses the high confidence problem as mentioned in Section \ref{sec:main:preprocess}. Obviously, we cannot simply combine multiple neural networks by concatenating their logits either if they are independently built by any other training methods.

\smalltitle{Logit scale problem}
The problem here is the fact that logits of different models can be in arbitrary scales and hence cannot simply be concatenated into a unified logit. Thus, we have to normalize logits in different scales, which we call the \textit{logit scale problem}. To illustrate, consider a composite task $Q$ consisting of $H_1 = \{cat, fox\}$ and $H_2 = \{dog, wolf\}$ in Figure \ref{fig:logitprob}. Also, let us say we have built two expert models for $H_1$ and $H_2$, which can be combined to the unified task-specific model $M(Q)$. Even if both $M(H_1)$ and $M(H_2)$ are properly trained to be not so confident about out-of-distribution samples as shown in Figure \ref{fig:logitprob:a}, the unified model $M(Q)$ can make a wrong inference when their logits are simply concatenated without considering their scales as shown in Figure \ref{fig:logitprob:b}. This can happen because the standard KD loss $\mathcal{L}_{KD}$ as well as our $\mathcal{L}_{soft}$ loss only focuses on trying to mimic the soft target, that is, a vector already normalized by the softlmax function such as $\sigma(\mathbf{t}_{H_i}/T)$ and $\sigma(\mathbf{s}_{H_i}/T)$. Although this soft target indeed helps to extract the knowledge of a teacher model, its logit value can end up in an arbitrary range after the distillation process. Consequently, the output logits of experts can be in quite different scales. Once we lose the overall scale information on logits with respect to the whole of classes, there is no way of restoring it unless we turn to the oracle network.

In order to address this logit scale problem, we define the loss $\mathcal{L}_{scale}$ as a regularization term of $\L_{CKD}$, aiming to transfer the scale information as well during the conditional distillation process. Unlike $\L_{soft}$, $\mathcal{L}_{scale}$ attempts to match hard targets by:
\begin{equation} \label{hard loss}
\mathcal{L}_{scale} = \parallel \mathbf{t}_{H_i}-\mathbf{s}_{H_i} \parallel_1.
\end{equation}
The reason why we use $L^1$-loss here is that we have to convey the overall scale information of a logit by $\mathcal{L}_{scale}$, rather than carrying its individual values themselves. Usually, $L^1$-loss is known to be more robust to outliers than $L^2$-loss, and this property can help us not to focus on exact logit values. Learning each expert by $\mathcal{L}_{scale}$ allows a logit of the expert to be in a larger scale than those of the other experts particularly for the samples relevant to the expert. In Section \ref{sec:experiments:service}, we show the effectiveness of $\mathcal{L}_{scale}$ by the experiments.

\section{Experimental Analysis} \label{sec:experiment}

\subsection{Evaluation Setting} \label{sec:experiment:setting}
\smalltitle{Datasets}
We conduct our experiments using two benchmark datasets for the image classification task, namely CIFAR-100 \cite{krizhevsky2009learning} and Tiny-ImageNet \cite{le2015tiny}. The CIFAR-100 dataset contains 50K training images and 10K test images for a total of 100 classes, and each class again belongs to one of 20 superclasses (e.g., flowers, people, etc.). We regard these 20 superclasses as primitive tasks.

Tiny-ImageNet contains 100K training images and 10K test images for 200 classes. For the primitive tasks, we refer to a semantic class hierarchy of ImageNet\footnote{http://image-net.org/explore} widely used in the literature of image classification. Considering the semantic tree, we group a few (from 3 to 10) classes at the leaf level to be a primitive task such that classes belonging to the same primitive task are semantically similar, that is, having a common ancestor class at a low level of the tree. In our experiments, we randomly choose \add{six} of all the primitive tasks for both CIFAR-100 and Tiny-ImageNet.

\smalltitle{Model architectures}
As a base architecture for both generic and specialized models, we use \textit{wide residual networks} (WRNs) \cite{DBLP:conf/bmvc/ZagoruykoK16} as we can easily test various models with different sizes by adjusting the layer depth factor $l$ and widening factor $k$. More specifically, the basic structure of WRNs consists of four groups of convolutions, namely \texttt{conv1}, \texttt{conv2}, \texttt{conv3}, and \texttt{conv4}. Other than \texttt{conv1} with 16 channels, the number of channels can be controlled by a common factor $k$ such that \texttt{conv}$i$ has $16 \times 2^{(i-2)} \times k$ channels. In this paper, we extend this basic WRN structure so that $k$ is further divided into two different factors, namely $k_c$, and $k_s$, such that the number of channels in either \texttt{conv2} or \texttt{conv3} is commonly determined by $k_c$, that is, $16 \times k_c$ for \texttt{conv2} and  $32 \times k_c$ for \texttt{conv3}, whereas that of \texttt{conv4} is independently controlled by $k_s$, that is, $64 \times k_s$. The reason for using this fine-grained WRN structure is to further reduce the number of channels of solely the final convolution group (i.e., \texttt{conv4}) for specialized models. Naturally extending the notation of a basic WRN, which is WRN-$l$-$k$, we denote our fine-grained version of WRNs as WRN-$l$-$(k_c, k_s)$.

For our branched architecture (see Figure \ref{fig:branched architecture}) of task-specific models returned from PoE, we set \texttt{conv4} solely to be an expert component, and the other convolution groups to be the library part in the experiment. Also, for each queried task $Q$, we denote this branched architecture as WRN-$l$-$(k_c, [k_{s_{1 \cdots n(Q)}}]^T)$, where a column vector $[k_{s_{1 \cdots n(Q)}}]^T$ indicates a series of $n(Q)$ widening factors, each of which corresponds to one of the $n(Q)$ branching experts (i.e., $n(Q)$ \texttt{conv4} blocks). It is noteworthy that $n(Q)$ \texttt{conv4} blocks, each having $64 \times k_s$ channels, are not exactly the same as a single \texttt{conv4} block with $n(Q)\times 64 \times k_s$ channels. To be shown in Table \ref{tab:merge test}, these branched blocks are also effective on reducing the model size as the number of parameters of $n(Q)$ \texttt{conv4} blocks is $n(Q)$ times larger than that of \texttt{conv4} while a single \texttt{conv4} block with $n(Q)$ times more channels has $n(Q)^2$ times more parameters.

\smalltitle{Implementation and training details} All algorithms\footnote{https://github.com/bigdata-inha/Pool-of-Experts-code} were implemented using PyTorch \cite{PaszkeGMLBCKLGA19} and evaluated on a machine with an \add{NVIDIA Quadro RTX 6000 and Intel Core Xeon Gold 5122}. When training all the models, we use a stochastic gradient descent (SGD) \cite{kiefer1952stochastic} with 0.9 momentum and the weight decay of $L^2$ regularization was fixed to $5\times10^{-4}$. The batch size of all networks was set to 512 \add{and $\alpha$ is always fixed to be $0.3$}.

\subsection{Experiments on Model Specialization} \label{sec:experiment:special}

\smalltitle{Oracle models and library extraction} For the oracle models that would be sufficiently large in practice, we use WRN-40-(4, 4) for CIFAR-100, and WRN-16-(10, 10) for Tiny-ImageNet, both of which are trained from scratch. The student models used in the library extraction are WRN-16-(1, 1) and WRN-16-(2, 2) for CIFAR-100 and Tiny-ImageNet, respectively, which are distilled from oracles by Eq. (\ref{KD loss}). As shown in Table \ref{tab:Oracle and library}, the resulting library models themselves are less accurate than oracles due to their lightweight architectures. We take blocks from \texttt{conv1} to \texttt{conv3} of these library models to be the library component shared by all task-specific models.

%%% Teacher and student for library
\begin{table}[t]
\begin{center}
    \scriptsize
    \begin{tabular}{|c|c||c|c|c|}
        % CIFAR100
        \hline
        \multicolumn{5}{c}{CIFAR-100} \\
        \hline
        \multicolumn{2}{|c||}{\textbf{Model}} &  \textbf{Acc.} & \textbf{FLOPs} & \textbf{Params}\\
        \hline
        Oracle (teacher) &  WRN-40-(4, 4) &  76.70 & 1.30B & 8.97M\\
        \textit{Library model} (student) & WRN-16-(1, 1) &  63.84 & 0.03B & 0.18M\\
        \hline
        % TinyImageNet
        \hline
        \multicolumn{5}{c}{Tiny-ImageNet} \\
        \hline
        \multicolumn{2}{|c||}{\textbf{Model}} &  \textbf{Acc.} & \textbf{FLOPs} & \textbf{Params}\\
        \hline
        Oracle (teacher) &  WRN-16-(10, 10) &  64.49 & 2.42B & 17.24M\\
        \textit{Library model} (student) & WRN-16-(2, 2) & 56.96 & 0.10B & 0.72M\\
        \hline
    \end{tabular}
    \vspace{1mm}
    \caption{Accuracy and model sizes of oracles and student models for \textit{library}, both of which are generic}
    \label{tab:Oracle and library}
    \vspace{-8mm}
\end{center}
\end{table}

\smalltitle{Experts extraction}
The architecture for an expert is set to WRN-16-(1, 0.25) for CIFAR-100 and WRN-16-(2, 0.25) for Tiny-ImageNet. As depicted in Figure \ref{fig:branched architecture}, all the experts share the library component (i.e., \texttt{conv1}, \texttt{conv2}, and \texttt{conv3}) that is fixed during the process of our CKD method. To extract each expert, CKD updates only \texttt{conv4} whose size is now reduced by a factor of 4 for CIFAR-100 and by a factor of 8 for Tiny-ImageNet, compared to the number of channels in \texttt{conv4} of its corresponding library model.

% In the experiment, $M(H_i)$'s architecture for each $H_i$ in CIFAR-100 used WRN-16-(1, 1, 0.25). Until conv2 block of $M(H_i)$ is the \textit{library}, and the other back is each expert $E_i$ corresponding to $H_i$. In other words, WRN-16-(1, 1, X) corresponds to \textit{library} and WRN-16-(X, X, 0.25) corresponds to $E_i$. In the TinyImageNet experiment, WRN-16-(2, 2, 0.25) was used as an architecture of $M(H_i)$ for each $H_i$. Similar to CIFAR-100, WRN-16-(2, 2, X) corresponds to the \textit{library}, and each $E_i$ corresponds to WRN-16-(X, X, 0.25).

\smalltitle{Model specialization methods} We consider the following baselines that can also build expert models with the same architecture, and compare their performance with our CKD method.
\begin{itemize}
    \item \smalltitle{KD} This is standard KD, where the entire knowledge of oracle is distilled to the lightweight architecture for expert using all the original data, thereby yielding a generic model.
    \item \smalltitle{Scratch} Without the help of oracle, this method trains the same architecture from scratch using only the task-specific dataset corresponding to each primitive task.
    \item \smalltitle{Transfer} Similar to our CKD method, this updates only the expert component (i.e., \texttt{conv4}) by performing transfer learning from the library component (i.e., from \texttt{conv1} to \texttt{conv3}) on the task-specific data for each primitive task.
    \item \smalltitle{CKD (ours)} This is our conditional knowledge distillation method using all the original training data like KD.
\end{itemize}
In order to fairly evaluate the effectiveness of each method for training a specialized model, we do not measure the overall accuracy for generic models with respect to the whole set of classes. That would be always lower than the accuracy of specialized models. Instead, we locally compare only the probability values corresponding to the target primitive task, and take the label with a maximum probability within the task as a prediction. We call the accuracy of generic models measured in this way \textit{task-specific accuracy} for the rest of the section.

%The above baseline and Oracle are largely divided into Generic models and Specialized models. For a fair experiment, we did not measure the accuracy of the Generic model for the entire tasks, but for the primitive task. In other words, accuracy was measured by comparing only the value corresponding to the primitive task among the output probabilities of the entire tasks. We call the accuracy of Generic model measured in this way in the rest of the experiment \textit{target task accuracy}.

%%% specialized test table
\begin{table}[t]
\begin{center}
    \scriptsize
    \begin{tabular}{|c|c|c||c|c|c|}
        % CIFAR100
        \hline
        \multicolumn{6}{c}{CIFAR-100} \\
        \hline
        \textbf{Method} & \textbf{Type} & \textbf{Architecture} & \textbf{Acc.} & \textbf{FLOPs} & \textbf{Params}\\
        \hline
        Oracle & \multirow{2}{*}{generic} & WRN-40-(4, 4) &  85.80$_{\pm11.9}$ & 1.30B & 8.97M\\
        \cline{1-1}\cline{3-6}
        \cline{1-1}\cline{3-6}
        KD & & \multirow{4}{*}{WRN-16-(1, 0.25)} &  62.50$_{\pm16.5}$ & \multirow{3}{*}{0.02B} & \multirow{3}{*}{0.06M}\\
        \cline{1-2}\cline{4-4}
        Scratch & \multirow{3}{*}{special} &  &  74.20$_{\pm14.7}$ &  & \\
        \cline{1-1}\cline{4-4}
        Transfer & & & 78.33$_{\pm12.1}$ & ($\times~ \frac{1}{65} $) & ($\times~\frac{1}{150} $) \\
        \cline{1-1}\cline{4-4}
        CKD (ours) & & & \textbf{82.40$_{\pm11.8}$} &  & \\
        \hline
        % TinyImageNet
        \hline
        \multicolumn{6}{c}{Tiny-ImageNet} \\
        \hline
        \textbf{Method} & \textbf{Type} & \textbf{Architecture} & \textbf{Acc.} & \textbf{FLOPs} & \textbf{Params}\\
        \hline
        Oracle &  \multirow{2}{*}{generic} & WRN-16-(10, 10) &  79.68$_{\pm5.4}$ & 2.42B & 17.24M\\
        \cline{1-1}\cline{3-6}
        \cline{1-1}\cline{3-6}
        KD & & \multirow{4}{*}{WRN-16-(2, 0.25)} &  57.62$_{\pm6.7}$ & \multirow{3}{*}{0.07B} & 0.19M\\
        \cline{1-2}\cline{4-4}\cline{6-6}
        Scratch & \multirow{3}{*}{special} &  &  66.10$_{\pm5.3}$ & & 0.18M\\
        \cline{1-1}\cline{4-4}
        Transfer & & & 74.21$_{\pm5.1}$ &  ($\times~\frac{1}{35} $) & ($\times~\frac{1}{96} $)\\
        \cline{1-1}\cline{4-4}
        CKD (ours) & & &  \textbf{78.72$_{\pm5.6}$} &  & \\
        \hline
    \end{tabular}
    \vspace{1mm}
    \caption{\add{Comparison on the accuracy and model size of specialization methods}}
    \label{tab:specialized result}
\end{center}
\vspace{-4mm}
\end{table}

%%% High confidence graph
\begin{figure}[t!]
    \centering
    \vspace{-2mm}
    % Scratch
    \subfigure[Scratch]{\hspace{-1mm}\includegraphics[height=0.32\columnwidth]{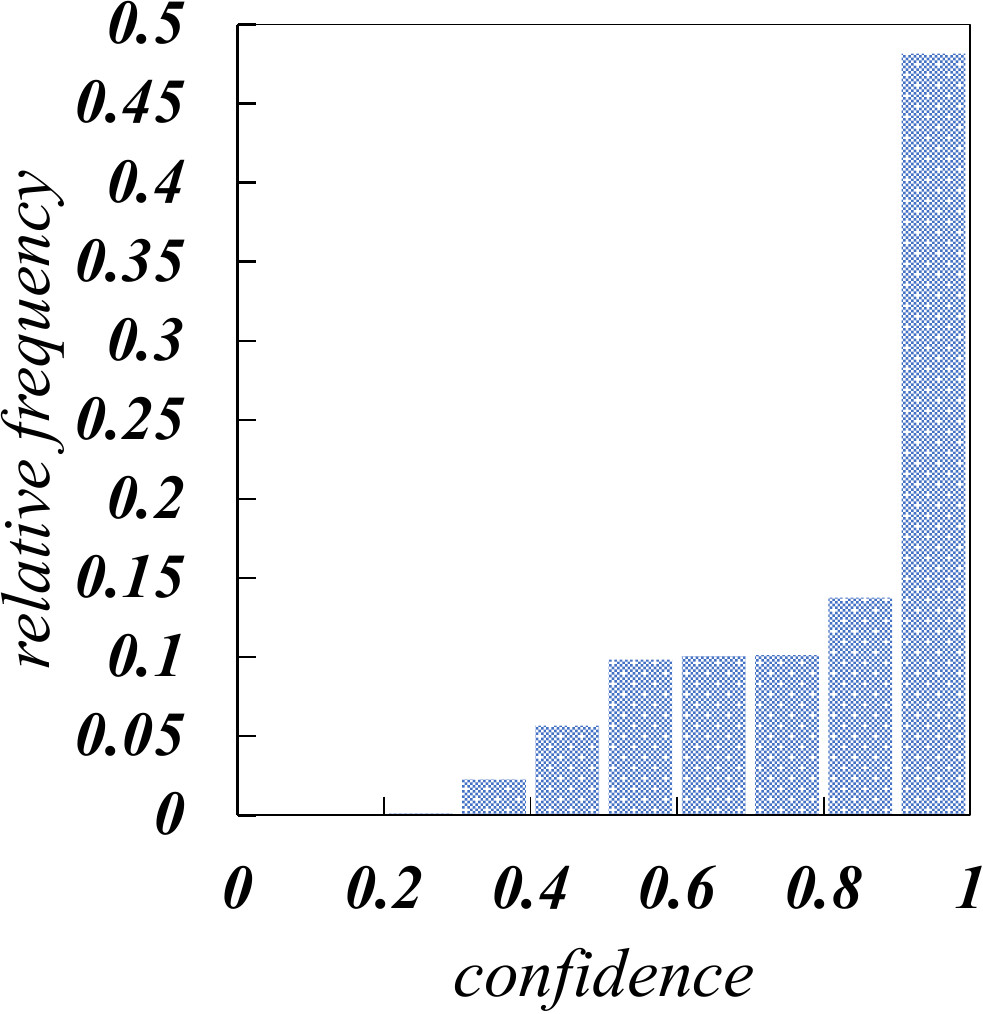}}
    % Transfer
    \subfigure[Transfer]{\includegraphics[height=0.32\columnwidth]{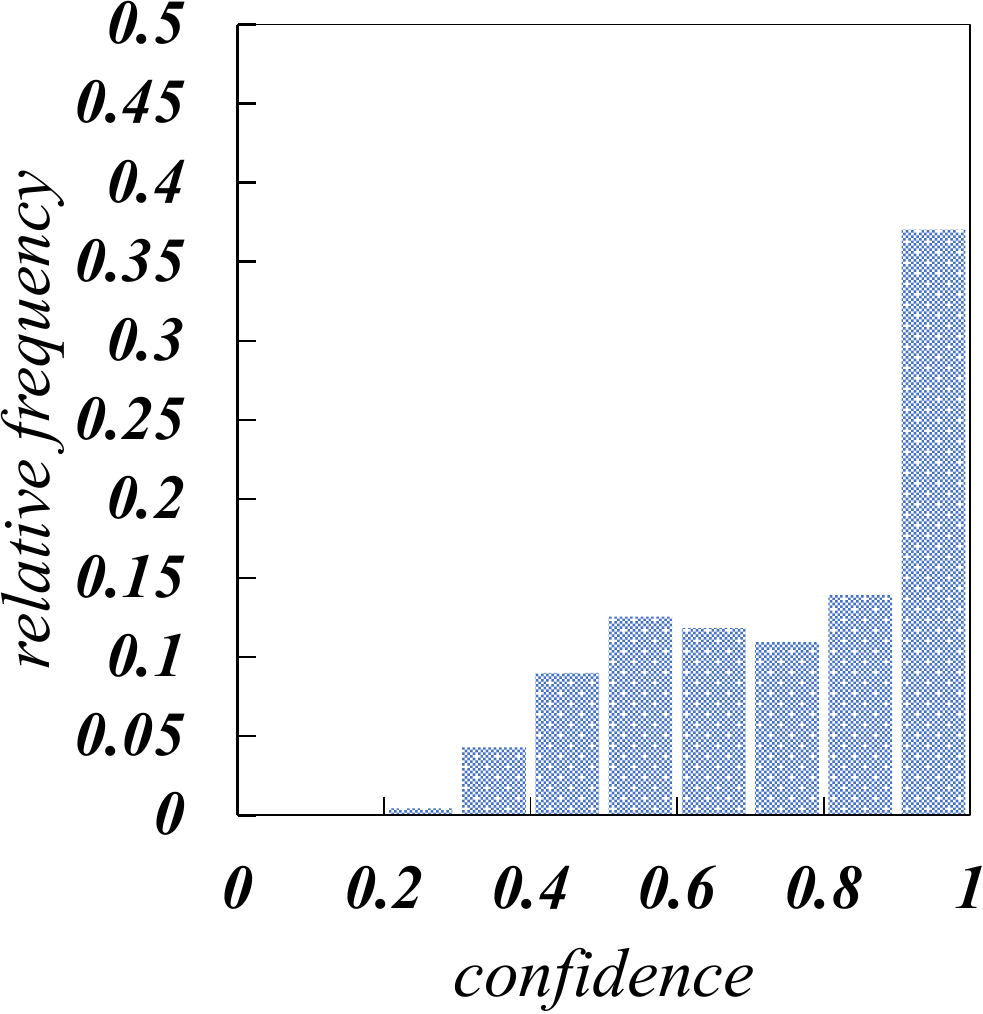}}
    % only soft
    \subfigure[CKD ($\mathcal{L}_{soft}$ or both)]{\includegraphics[height=0.32\columnwidth]{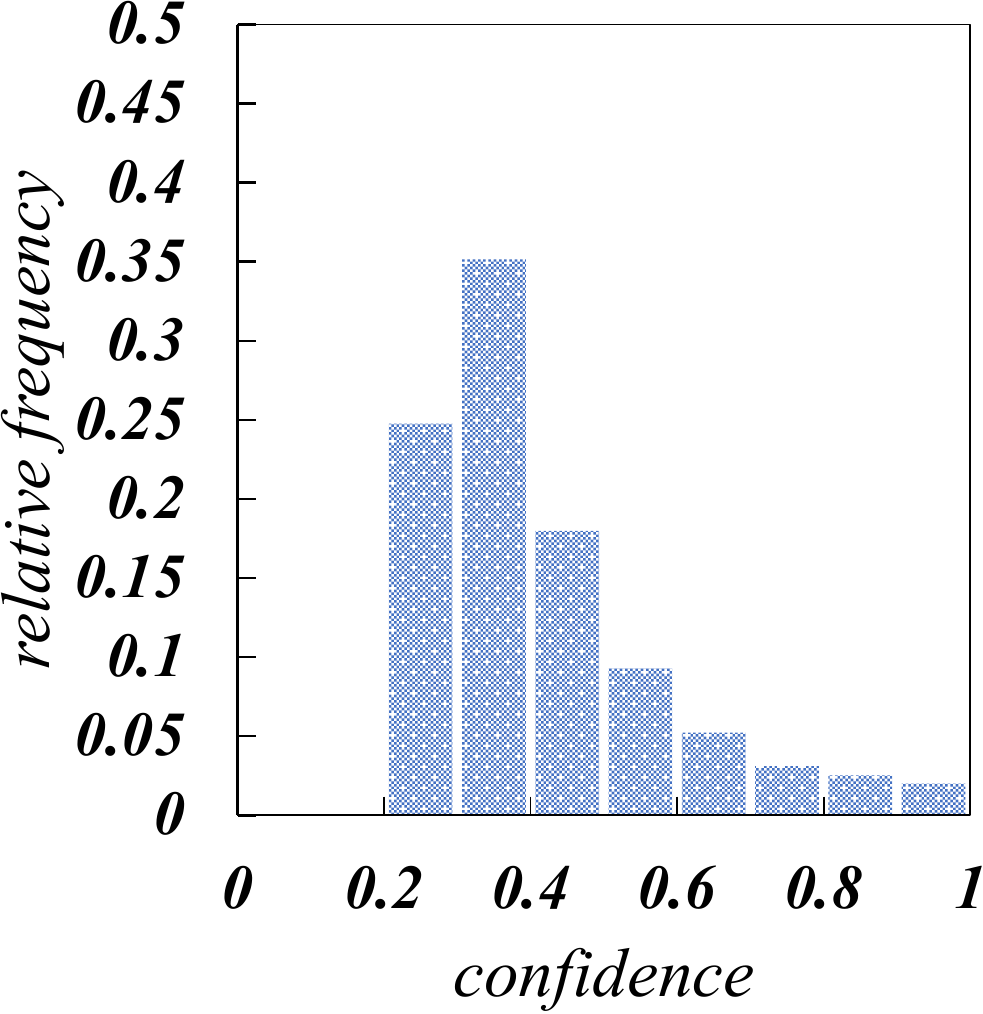}\hspace{0mm}}
    % only soft
    % \subfigure[CKD ($\mathcal{L}_{soft}$ and $\mathcal{L}_{scale}$)]{\hspace{2mm}\includegraphics[width=0.20\textwidth]{high confidence fig/soft and hard}\hspace{2mm}}
    \vspace{-4mm}
    \caption{Histograms on maximum confidence values (i.e., highest class probabilities) in models specialized by different methods for \textit{vehicles1} of CIFAR-100 as the primitive task}
    \label{fig:high confidence}
    \vspace{-4mm}
\end{figure}

% Specialized test result
\smalltitle{Accuracy and model size}
The experimental result on model specialization is summarized in Table \ref{tab:specialized result}. For the accuracy, we take the average value over \add{six} primitive tasks selected in a way described in Section \ref{sec:experiment:setting}. As mentioned above, we measure the task-specific accuracy for generic models and the normal accuracy for specialized models. In terms of model sizes, specialized models use two orders of magnitude less parameters than their oracle in both datasets, and hence oracle maintains the best accuracy for all the primitive tasks. 

Among the specialization methods, our CKD method achieves the best accuracy pretty close to the level of oracle's, despite its tiny target models. This result confirms that CKD successfully extracts only the essential specialized knowledge from oracle. In contrast, KD clearly fails to compress the entire knowledge to fit in these small-sized models to the point that its task-specific accuracy is always the lowest among all the methods.
\add{Furthermore, as observed that CKD performs better than Transfer with clear margins, knowledge distillation can surely be regarded as a more effective way of acquiring the oracle's knowledge than learning it directly from data samples like Transfer in the sense that both CKD and Transfer exploit the same shared component, i.e., library.}

% The average Accuracy and FLOPs of specialized models learned through baseline method for the 5 primitive tasks are shown in the Table \ref{tab:specialized result} above. First of all, the KD showed significantly lower accuracy compared the CKD. This shows that when a model is needed that classifies only a few target tasks of Oracle, using KD to distill Oracle's entire knowledge shows that it is inefficient to have low accuracy by wasting the model's capacity. It can also be seen that CKD has a higher accuracy than the rest of the specialized methods as well as KD.

\smalltitle{Confidence analysis for specialized models}
% high confidence description
Let us now see how well CKD trains a properly confident expert by using the $\mathcal{L}_{soft}$ loss. To this end, we collect the highest probabilities resulting from each specialized model when we make a prediction using \textit{out-of-distribution} samples, that is, images not belonging to any class of the corresponding primitive task. Obviously, any prediction made in this way cannot be correct simply because each model does not include the output label for the correct class to be predicted. 

Figure \ref{fig:high confidence} presents an example to see how those \textit{highest confidence values} are distributed, all of which are from out-of-distribution samples. It is clearly observed that Transfer as well as Scratch is too confident about out-of-distribution samples in that a confidence value larger than $0.9$ appears most frequently. Thus, as mentioned in the introduction, even if we perform transfer learning from oracle to obtain a specialized model, the resulting model could be overconfident about samples irrelevant to it. As opposed to that, in the models specialized by CKD, the most frequent confidence value lies in the range from $0.3$ to $0.4$, implying not too confident about classes they do not know, when using either only $\mathcal{L}_{soft}$ or $\mathcal{L}_{CKD}$ as a whole. This confirms the fact that $\mathcal{L}_{soft}$ effectively informs the target model about unknown classes in the process of distillation.
\subsection{Experiments on Model Consolidation} \label{sec:experiments:service}

\smalltitle{Model consolidation methods} The performance test in the service phase lies in the main objective of \ours to examine how fast and efficiently \ours generates the \textit{task-specific model} $M(Q)$ for a queried \textit{composite task} $Q$. For the experiment of the service phase, we consider the following methods:

\begin{itemize}
    \item \textbf{Scratch}, \textbf{Transfer}, and \textbf{CKD (ours).}\hspace{1mm} These three training methods are basically the same as they are in the specialization test, but now we train the task-specific model $M(Q)$ using its task-specific dataset for a given composite task $Q$ instead of each primitive task.
    \item \textbf{SD}/\textbf{UHC} $+$ \textbf{Scratch}/\textbf{CKD.}\hspace{1mm} 
    As another group of methods for the service phase, we consider neural network merging with pre-built primitive models. Both SD and UHC are the methods of merging individually trained classifiers as introduced in \cite{VongkulbhisalVS19}. SD is a naive extension of standard distillation and used as a baseline of UHC. In order to build $M(Q)$, for each primitive task $H_i$ constituting $Q$, we use the corresponding expert model, i.e., $M(H_i)$, as a teacher, and all the expert models are jointly distilled through the SD method or the UHC method into the predefined architecture for $M(Q)$. 
    Each primitive model $M(H_i)$ can be trained either from scratch or by our CKD method.
    \item \smalltitle{\ours(ours)} This is our final model consolidation method with no training process as described in Section \ref{sec:main:service}.
\end{itemize}
\vspace{-2mm}

\begin{table*}[t]
\begin{center}
    \scriptsize
    \begin{tabular}{|l|l|l||c|c|c|c|c|c|c|c|c|c|c|c|}
        % CIFAR-100
        \hline
        \multicolumn{15}{c}{CIFAR-100} \\
        \hline
        \multirow{3}{*}{\textbf{Method}} & \multirow{3}{*}{\textbf{Type}} & \multirow{3}{*}{\textbf{Architecture}} & \multicolumn{12}{c|}{Number of primitive tasks, $n(Q)$}\\
        \cline{4-15}
        & & & \multicolumn{3}{c|}{2} & \multicolumn{3}{c|}{3} & \multicolumn{3}{c|}{4} & \multicolumn{3}{c|}{5}\\
        \cline{4-15}
        & & & \textbf{Acc.} & \textbf{FLOPs} & \textbf{Params} & \textbf{Acc.} & \textbf{FLOPs} & \textbf{Params} & \textbf{Acc.} & \textbf{FLOPs} & \textbf{Params} & \textbf{Acc.} & \textbf{FLOPs} & \textbf{Params}\\
        \hline
        Oracle & \multirow{2}{*}{generic} & WRN-40-(4, 4) & 84.25$_{\pm7.3}$ & 1.30B & 8.97M & 82.94$_{\pm5.0}$ & 1.30B & 8.97M & 81.82$_{\pm3.3}$ & 1.30B & 8.97M & 80.82$_{\pm2.0}$ & 1.30B & 8.97M\\
        \cline{1-1}\cline{3-15}
        KD & & \multirow{7}{*}{WRN-16-} & 67.61$_{\pm10.0}$ & \multirow{7}{*}{0.02B} & \multirow{7}{*}{0.08M} & 71.29$_{\pm5.2}$ & \multirow{7}{*}{0.02B} & 0.13M & 72.32$_{\pm3.3}$ & \multirow{7}{*}{0.03B} & \multirow{7}{*}{0.18M} & 72.43$_{\pm1.7}$ & \multirow{7}{*}{0.03B} & 0.25M\\
        \cline{1-2}\cline{4-4}\cline{7-7}\cline{9-10}\cline{13-13}\cline{15-15}
        Scratch & \multirow{9}{*}{special} & & 72.65$_{\pm8.3}$ & & & 71.47$_{\pm5.3}$ & & \multirow{5}{*}{0.12M} & 70.97$_{\pm3.6}$ & & & 70.21$_{\pm2.0}$ & & \multirow{5}{*}{0.24M}\\
        \cline{1-1}\cline{4-4}\cline{7-7}\cline{10-10}\cline{13-13}
        Transfer & & & 77.82$_{\pm7.3}$ & & & 77.50$_{\pm5.4}$ & & & \underline{74.54$_{\pm3.7}$} & & & \underline{73.36$_{\pm2.2}$} & & \\
        \cline{1-1}\cline{4-4}\cline{7-7}\cline{10-10}\cline{13-13}
        SD+Scratch & & & 57.06$_{\pm9.3}$ & & & 48.60$_{\pm6.4}$ & & & 43.08$_{\pm4.1}$ & & & 39.15$_{\pm2.4}$ & & \\
        \cline{1-1}\cline{4-4}\cline{7-7}\cline{10-10}\cline{13-13}
        UHC+Scratch & & (1, 0.25 $\times n(Q)$) & 57.57$_{\pm10.3}$ & ($\times~ \frac{1}{65}$) & ($\times~ \frac{1}{112} $) & 49.73$_{\pm7.1}$ & ($\times~ \frac{1}{65} $) & ($\times~ \frac{1}{75} $) & 44.49$_{\pm4.7}$ & ($\times~ \frac{1}{43} $) & ($\times~ \frac{1}{50} $) & 40.83$_{\pm2.7}$ & ($\times~ \frac{1}{43} $) & ($\times~ \frac{1}{37} $) \\
        \cline{1-1}\cline{4-4}\cline{7-7}\cline{10-10}\cline{13-13}
        SD+CKD & & & 73.94$_{\pm7.6}$ & & & 71.28$_{\pm5.7}$ & & & 69.46$_{\pm4.1}$ & & & 67.77$_{\pm1.9}$ & & \\
        \cline{1-1}\cline{4-4}\cline{7-7}\cline{10-10}\cline{13-13}
        UHC+CKD & & & 73.87$_{\pm7.9}$ & & & 71.56$_{\pm5.5}$ & & & 70.49$_{\pm3.8}$ & & & 68.84$_{\pm2.3}$ & & \\
        \cline{1-1}\cline{4-4}\cline{7-7}\cline{10-10}\cline{13-13}
        CKD(ours) & & & \underline{78.55$_{\pm7.5}$} & & & \textbf{77.00$_{\pm5.3}$} & & & \textbf{75.70$_{\pm3.5}$} & & & \textbf{74.27$_{\pm2.2}$} & & \\
        \cline{1-1}\cline{3-15}
        \multirow{2}{*}{PoE(ours)} & & WRN-16-(1, & \multirow{2}{*}{\textbf{79.03$_{\pm7.9}$}} & 0.02B & 0.07M & \multirow{2}{*}{\underline{76.41$_{\pm5.5}$}} & 0.02B & 0.08M & \multirow{2}{*}{74.18$_{\pm3.7}$} & 0.02B & 0.09M & \multirow{2}{*}{72.22$_{\pm2.2}$} & 0.02B & 0.10M\\
        & & [0.25$_{1\cdots n(Q)}$]$^T$) & & ($\times~ \frac{1}{65} $) & ($\times~ \frac{1}{128} $) & & ($\times~ \frac{1}{65} $) & ($\times~ \frac{1}{112} $) & & ($\times~ \frac{1}{65} $) & ($\times~ \frac{1}{100} $) & & ($\times~ \frac{1}{65} $) & ($\times~ \frac{1}{90} $) \\
        \hline
        % Tiny-ImageNet
        \hline
        \multicolumn{15}{c}{Tiny-ImageNet} \\
        \hline
        \multirow{3}{*}{\textbf{Method}} & \multirow{3}{*}{\textbf{Type}} & \multirow{3}{*}{\textbf{Architecture}} & \multicolumn{12}{c|}{Number of primitive tasks, $n(Q)$}\\
        \cline{4-15}
        & & & \multicolumn{3}{c|}{2} & \multicolumn{3}{c|}{3} & \multicolumn{3}{c|}{4} & \multicolumn{3}{c|}{5}\\
        \cline{4-15}
        & & & \textbf{Acc.} & \textbf{FLOPs} & \textbf{Params} & \textbf{Acc.} & \textbf{FLOPs} & \textbf{Params} & \textbf{Acc.} & \textbf{FLOPs} & \textbf{Params} & \textbf{Acc.} & \textbf{FLOPs} & \textbf{Params}\\
        \hline
        Oracle & \multirow{2}{*}{generic} & WRN-16-(10, 10) & 77.30$_{\pm4.0}$ & 2.42B & 17.24M & 75.65$_{\pm3.0}$ & 2.42B & 17.24M & 74.31$_{\pm2.1}$ & 2.42B & 17.24M & 73.18$_{\pm1.4}$ & 2.42B & 17.24M\\
        \cline{1-1}\cline{3-15}
        KD & & \multirow{7}{*}{WRN-16-} & 60.54$_{\pm6.0}$ & \multirow{7}{*}{0.07B} & 0.22M & 62.24$_{\pm4.2}$ & \multirow{7}{*}{0.07B} & 0.27M & 62.77$_{\pm2.4}$ & \multirow{7}{*}{0.08B} & 0.33M & 62.80$_{\pm1.5}$ & \multirow{7}{*}{0.08B} & 0.41M\\
        \cline{1-2}\cline{4-4}\cline{6-7}\cline{9-10}\cline{12-13}\cline{15-15}
        Scratch & \multirow{9}{*}{special} &  & 64.23$_{\pm3.8}$ & & \multirow{5}{*}{0.21M} & 63.65$_{\pm2.4}$ & & \multirow{5}{*}{0.26M} & 62.90$_{\pm2.1}$ & & \multirow{5}{*}{0.32M} & 63.02$_{\pm1.3}$ & & \multirow{5}{*}{0.39M}\\
        \cline{1-1}\cline{4-4}\cline{7-7}\cline{10-10}\cline{13-13}
        Transfer & & & 71.18$_{\pm4.4}$ & & & 70.14$_{\pm3.3}$ & & & 68.71$_{\pm2.9}$ & & & 67.49$_{\pm1.1}$ & & \\
        \cline{1-1}\cline{4-4}\cline{7-7}\cline{10-10}\cline{13-13}
        SD+Scratch & & & 48.38$_{\pm4.5}$ & & & 38.60$_{\pm3.2}$ & & & 33.39$_{\pm2.3}$ & & & 29.49$_{\pm1.2}$ & & \\
        \cline{1-1}\cline{4-4}\cline{7-7}\cline{10-10}\cline{13-13}
        UHC+Scratch & & (2, 0.25 $\times n(Q)$) & 51.81$_{\pm4.1}$ & ($\times~ \frac{1}{35} $) & ($\times~ \frac{1}{82} $) & 43.54$_{\pm2.9}$ & ($\times~ \frac{1}{35} $) & ($\times~ \frac{1}{66} $) & 38.42$_{\pm1.9}$ & ($\times~ \frac{1}{30} $) & ($\times~ \frac{1}{54} $) & 34.66$_{\pm1.6}$ & ($\times~ \frac{1}{30} $) & ($\times~ \frac{1}{44} $) \\
        \cline{1-1}\cline{4-4}\cline{7-7}\cline{10-10}\cline{13-13}
        SD+CKD & & & 64.44$_{\pm4.4}$ & & & 60.33$_{\pm3.8}$ & & & 57.42$_{\pm3.0}$ & & & 54.93$_{\pm2.0}$ & & \\
        \cline{1-1}\cline{4-4}\cline{7-7}\cline{10-10}\cline{13-13}
        UHC+CKD & & & 67.71$_{\pm3.3}$ & & & 65.43$_{\pm2.3}$ & & & 63.34$_{\pm1.6}$ & & & 61.85$_{\pm0.9}$ & & \\
        \cline{1-1}\cline{4-4}\cline{7-7}\cline{10-10}\cline{13-13}
        CKD(ours) & & & \underline{74.19$_{\pm4.6}$} & & & \textbf{72.90$_{\pm3.4}$} & & & \textbf{71.20$_{\pm2.9}$} & & & \textbf{70.14$_{\pm1.4}$} & & \\
        \cline{1-1}\cline{3-15}
        \multirow{2}{*}{PoE(ours)} & & WRN-16-(2, & \multirow{2}{*}{\textbf{74.68$_{\pm4.5}$}} & 0.07B & 0.20M & \multirow{2}{*}{\underline{71.84$_{\pm3.6}$}} & 0.07B & 0.22M & \multirow{2}{*}{\underline{69.59$_{\pm2.8}$}} & 0.07B & 0.23M & \multirow{2}{*}{\underline{67.71$_{\pm1.8}$}} & 0.07B & 0.25M\\
        & & [0.25$_{1\cdots n(Q)}$]$^T$) & & ($\times~ \frac{1}{35} $) & ($\times~ \frac{1}{86} $) & & ($\times~ \frac{1}{35} $) & ($\times~ \frac{1}{78} $) & & ($\times~ \frac{1}{35} $) & ($\times~ \frac{1}{75} $) & & ($\times~ \frac{1}{35} $) & ($\times~ \frac{1}{69} $) \\
        \hline
    \end{tabular}
    \vspace{1mm}
    \caption{\add{Comparison on the accuracy and size of task-specific models built by the compared methods, where the task-specific accuracy is presented for generic models}}
    \label{tab:merge test}    
\end{center}
\vspace{-8mm}
\end{table*}

% Model consolidation description
\smalltitle{Accuracy and model size} The experimental result on model consolidation is summarized in Table \ref{tab:merge test}. We measure the task-specific accuracy for generic models as in the preprocessing phase, and vary the number of primitive tasks in $Q$ from 2 to 5. For each accuracy value, we report the average over all the combinations of as many primitive tasks used in $Q$. 

In terms of the overall accuracy, it is surprisingly observed that \ours beats most of the training methods except for CKD even though \ours does not involve any training process. Our CKD method still shows almost always the highest accuracy as it is in the specialization test. Another interesting result is that SD + CKD and UHC + CKD show much higher accuracy than SD + Scratch and UHC + Scratch, respectively. This implies that the resulting model merged by SD or UHC is highly dependent on how the expert teacher models are trained. When SD and UHC work with CKD, they sort of take advantage of the \textit{composable} form of specialized knowledge extracted from oracle by our CKD method. In contrast, when they attempt to merge totally independently trained neural networks (i.e., SD + Scratch and UHC + Scratch), their accuracy gets significantly lower probably due to the problem of overconfident experts or the logit scale problem. 

Despite no further training phase, PoE can effectively consolidate specialized knowledge by addressing both of the problems above to the point that it shows almost always the highest accuracy except for CKD. This result is even more remarkable in the sense that every task-specific model built by PoE carries less parameters than the other trained models thanks to our proposed branched architecture as mentioned in Section \ref{sec:experiment:setting}.

\smalltitle{\add{Volume of the entire PoE framework}} 
\add{Another strength of PoE lies in the fact that the entire PoE framework requires only a small amount of  storage as presented in Table \ref{tab:Volumes}. As mentioned in the introduction, if we train and compress all $2^n$ specialized models in the preprocessing phase, the total volume is estimated to be extremely large such as at least 1198.40 terabytes for 34 primitive tasks of Tiny-ImageNet. In contrast, our PoE framework is managed to use only a few megabytes to store all the primitive experts together with one library component, which is about 20-30 times smaller than the volumes of oracles themselves.}

%%% Volumes on the entire PoE framework
\begin{table}[htb]
\begin{center}
    \scriptsize
    \begin{tabular}{|c||c|c|c|c|c|}
        % CIFAR100
        \hline
        \multicolumn{6}{c}{Volumes} \\
        \hline
        \multirow{2}{*}{Dataset} & \multirow{2}{*}{Oracle} & \multicolumn{3}{c|}{PoE} & All specialized\\
        \cline{3-5}
        & & Library & Expert & All &  (estimation) \\
        \cline{1-6}
        CIFAR100 & 34.3MB & 177KB & 54.3KB & \textbf{1.23MB} & $\geq$ 54.30GB\\
        \cline{1-6}
        Tiny-ImageNet & 65.8MB & 656KB & 74.9KB & \textbf{3.20MB} & $\geq$ 1198.40TB\\
        \hline
    \end{tabular}
    \vspace{1mm}
    \caption{\add{Volumes of the entire PoE framework}}
    \label{tab:Volumes}
    \vspace{-8mm}
\end{center}
\end{table}

\begin{figure*}[t]
    \changecolor{blue} %% to be deleted in the CRC version
	\begin{minipage}[h]{\columnwidth}
        \centering
        % CIFAR-100, n(Q)=5
        \subfigure[CIFAR-100, $n(Q)=5$]{\includegraphics[width=0.46\columnwidth]{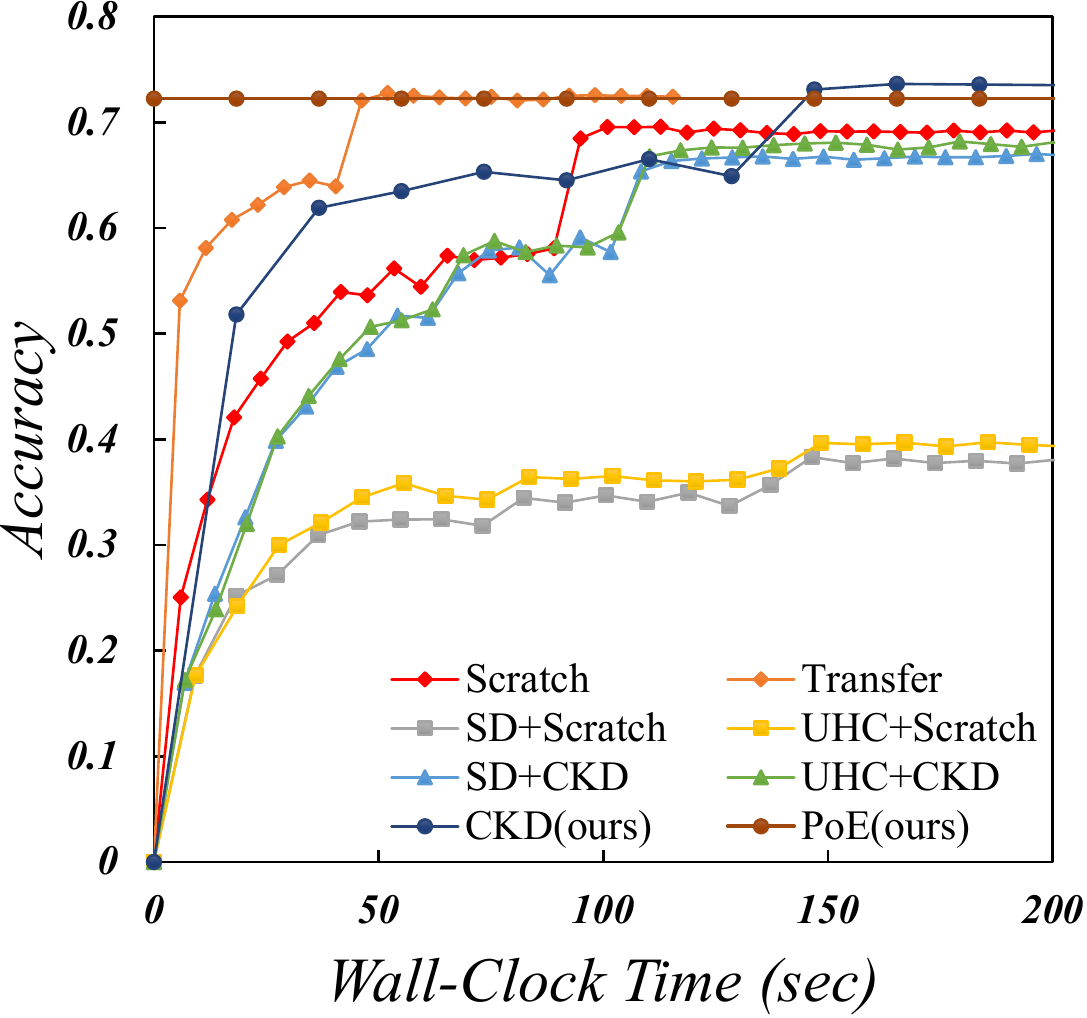}}
        % Tiny-ImageNet, n(Q)=5
        \subfigure[Tiny-ImageNet, $n(Q)=5$]{\includegraphics[width=0.46\columnwidth]{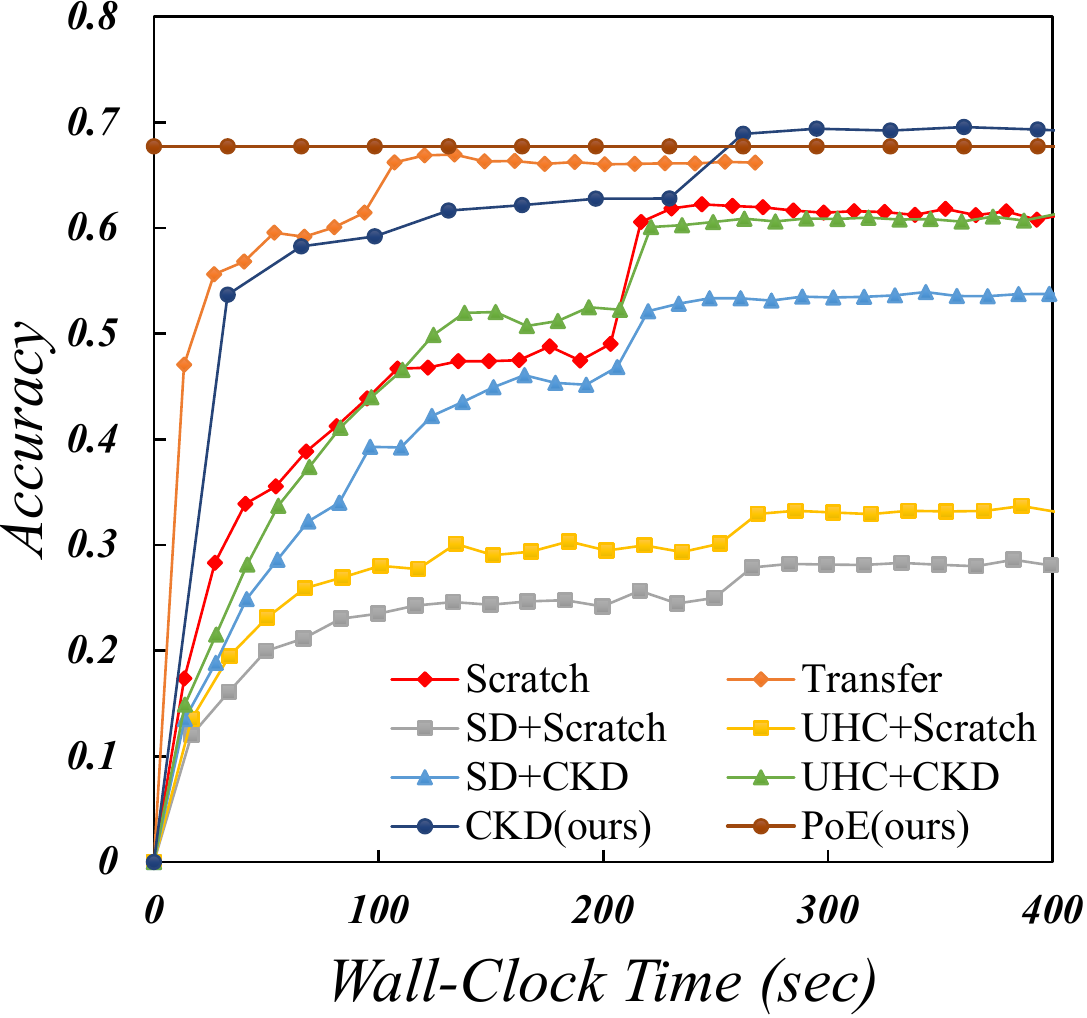}}
    \vspace{-4mm}
        \caption{\add{Learning curve of each training method in the service phase, where time is measured every 5 epochs}}
        \label{fig:5merge lr curve}
	\end{minipage}
    \hspace{4mm}
	\begin{minipage}[h]{\columnwidth}
        \centering
        % CIFAR-100
        \subfigure[CIFAR-100]{\includegraphics[width=0.45\columnwidth]{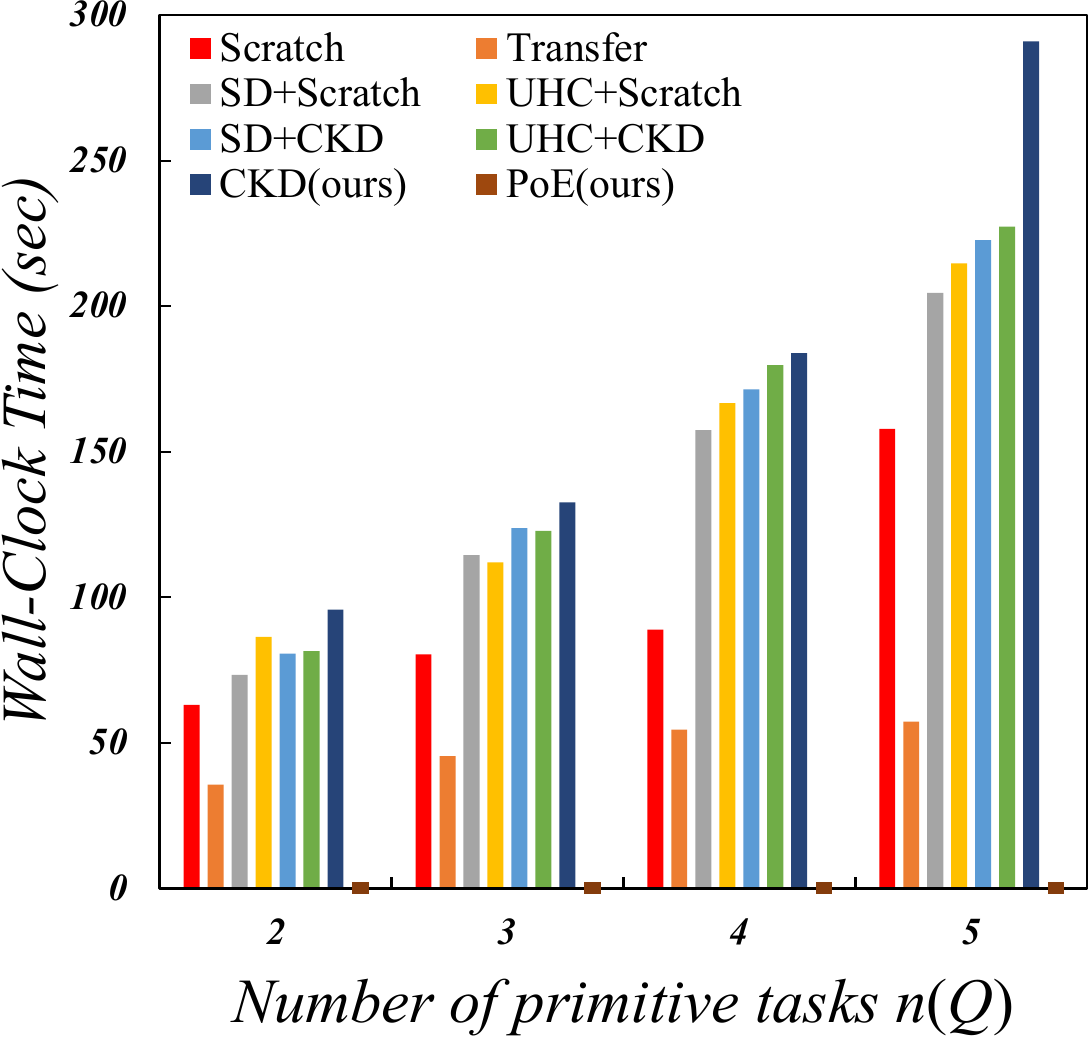}}
        % Tiny-ImageNet
        \subfigure[Tiny-ImageNet]{\includegraphics[width=0.45\columnwidth]{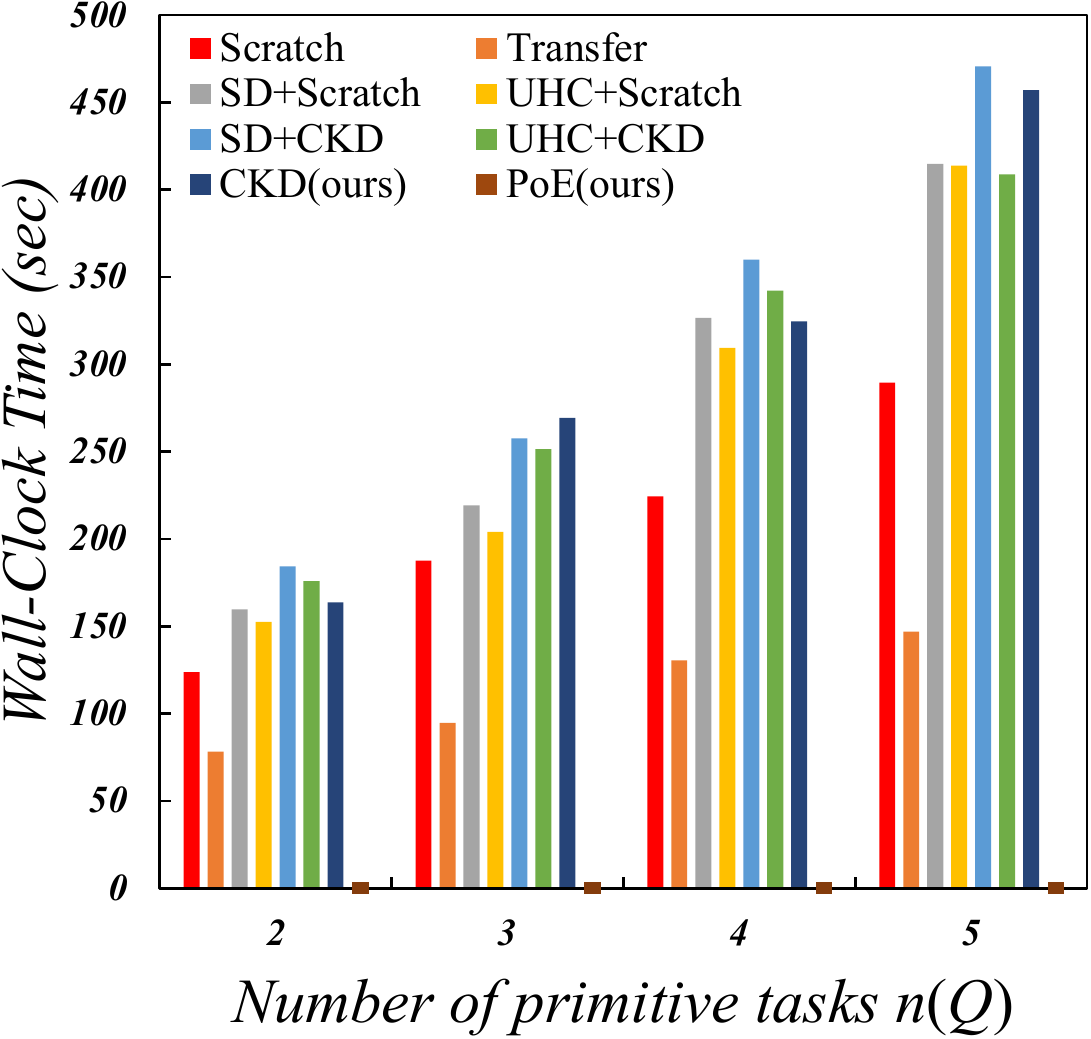}}
    \vspace{-4mm}
        \caption{\add{Training time for each task-specific model $M(Q)$ when varying $n(Q)$ from $2$ to $5$}}
        \label{fig:merge time}
	\end{minipage}
	\vspace{-4mm}
\end{figure*}

\begin{table}[htb]
\begin{center}
    \scriptsize
    \begin{tabular}{|c||c|c|c|c|}
        % CIFAR100
        \hline
        \multicolumn{5}{c}{CIFAR100} \\
        \hline
        \multirow{2}{*}{Method} & \multicolumn{4}{c|}{Number of primitive tasks, $n(Q)$}\\
        \cline{2-5}
        & 2 & 3 & 4 & 5\\
        \cline{1-5}
        $\mathcal{L}_{soft}$ only & 78.17$_{\pm8.1}$ & 75.61$_{\pm5.7}$ & 73.53$_{\pm3.9}$ & 71.76$_{\pm2.3}$\\
        \cline{1-5}
        $\mathcal{L}_{scale}$ only & 71.46$_{\pm8.5}$ & 68.44$_{\pm5.7}$ & 65.85$_{\pm3.6}$ & 63.59$_{\pm1.9}$\\
        \cline{1-5}
        $\mathcal{L}_{soft} + \L_{scale}$ & \textbf{79.03$_{\pm7.9}$} & \textbf{76.41$_{\pm5.5}$} & \textbf{74.18$_{\pm3.7}$} & \textbf{72.22$_{\pm2.2}$}\\
        \hline
        % TinyImageNet
        \hline
        \multicolumn{5}{c}{Tiny-ImageNet} \\
        \hline
        \multirow{2}{*}{Method} & \multicolumn{4}{c|}{Number of primitive tasks, $n(Q)$}\\
        \cline{2-5}
        & 2 & 3 & 4 & 5\\
        \cline{1-5}
        $\mathcal{L}_{soft}$ only & 73.25$_{\pm5.0}$ & 69.55$_{\pm3.8}$ & 66.72$_{\pm2.9}$ & 64.44$_{\pm2.0}$\\
        \cline{1-5}
        $\mathcal{L}_{scale}$ only & 68.95$_{\pm5.7}$ & 66.12$_{\pm4.4}$ & 63.90$_{\pm3.2}$ & 62.08$_{\pm2.1}$\\
        \cline{1-5}
        $\mathcal{L}_{soft} + \L_{scale}$ & \textbf{74.68$_{\pm4.5}$} & \textbf{71.84$_{\pm3.6}$} & \textbf{69.59$_{\pm2.8}$} & \textbf{67.71$_{\pm1.8}$}\\
        \hline
    \end{tabular}
    \vspace{1mm}
    \caption{\add{\add{Comparison on the average accuracies when using $\L_{soft}$, $\L_{scale}$, or both}}}
    \label{tab:soft and scale}
    \vspace{-8mm}
\end{center}
\end{table}

\smalltitle{Knowledge consolidation time}
In order to examine how long it takes for each of the other training methods to obtain its final task-specific model, we show the learning curve of each method in the case of $n(Q)=5$ in Figure \ref{fig:5merge lr curve}. Each learning curve is drawn by measuring time and accuracy every 5 epochs. We can observe that all the training based methods take from 50 seconds to 150 seconds for CIFAR-100 and from 100 seconds to 250 seconds for Tiny-ImageNet to get to their best result. For \ours, it is shown that it takes no time to reach the highest accuracy due to its train-free property. Considering a practical AIaaS system that has to process multiple queries with a bigger oracle model, a few minutes of time per query is prohibitive for the scalability of the system. Moreover, as shown in Figure \ref{fig:merge time}, when each queried composite-task consists of more and more primitive tasks, the average training time of all the methods \add{to achieve their best accuracy} drastically increases except for \ours. Thus, when a user requests a variety of composite tasks, only \ours is suitable for a realtime service.

% Additional misclassification description
\smalltitle{\add{Comparison between $\L_{soft}$ and $\L_{scale}$}}
\add{Lastly, we conduct an empirical study to examine the difference of training with $\L_{soft}$ and $\L_{scale}$. To this end, we additionally build two different versions of PoE, either of which is trained by only $\L_{soft}$ or $\L_{scale}$, and compare their resulting accuracies to those of our original version of PoE using both $\L_{soft}$ and $\L_{scale}$. As clearly observed in Table \ref{tab:soft and scale}, $\L_{scale}$ cannot effectively distill the oracle's knowledge alone, but essentially needs the help of $\L_{soft}$. At the same time, however, $\L_{scale}$ seems to be still helpful for addressing the logit scale problem as PoE performs always better with clear margins when using both $\L_{soft}$ and $\L_{scale}$ than using only $\L_{soft}$. }

\section{Conclusions} \label{sec:con}
The problem of realtime querying specialized knowledge over a massive pretrained neural network has never been dealt with in both the machine learning community and the database community. In this paper, through Pool of Experts (PoE), we have shown a new perspective that a large and generic neural network can be seen as a \emph{database} that can be preprocessed and re-organized so as to query any partial information like a tiny specialized neural network. In image classification, our proposal was highly successful as PoE can immediately synthesize any task-specific model accurate and small enough with no training at all. As a future work, we hope to see our new perspective on neural networks can be extended to more complicated machine learning problems like object detection and language translation. 
\vspace{-4mm}
\begin{acks}
\footnotesize
This work was supported in part by the National Research Foundation of Korea (NRF) grant funded by the Korea government (MSIT) (NRF-2018R1D1A1B07049934) and in part by Institute of Information \& communications Technology Planning \& Evaluation (IITP) grants funded by the Korea government (MSIT) (2019-0-00240, 2019-0-00064, and 2020-0-01389, Artificial Intelligence Convergence Research Center(Inha University)). 

Special thanks to Yeong-Hwa Jin and Jong-Yeong Kim for helpful discussions.
\end{acks}

%\newpage
\bibliographystyle{ACM-Reference-Format}
\balance
\bibliography{poebib}   % name your BibTeX data base

\appendix








\end{document}